\documentclass[onecolumn]{aastex6}
\usepackage{multirow} 
\usepackage{booktabs} 
\newcounter{magicrownumbers}

\slugcomment{Accepted for publication in the Astrophysical Journal}  
\shorttitle{SS Aur \& TU Men  
}
\shortauthors{Godon \& Sion}
\watermark{Astroph Version}
\setwatermarkfontsize{0.9in}

\begin{document}

\title{{\bf 
White Dwarf Photospheric Abundances in Cataclysmic Variables: \\
I. SS Aurigae and TU Mensae 
}
\footnote{Based on observations made with the NASA/ESA Hubble Space Telescope,
obtained from the data archive at the Space Telescope Science Institute.
STScI is operated by the Association of University for Research in 
Astronomy, Inc. under NASA contract NAS 5-26555.}
}

\author{Patrick Godon\altaffilmark{1,2} }

\and

\author{Edward M. Sion\altaffilmark{1} } 

\email{patrick.godon@villanova.edu}

\altaffiltext{1}{Department of Astrophysics \& Planetary Science, 
Villanova University, Villanova, PA 19085, USA}
\altaffiltext{2}{Henry A. Rowland Department of Physics \& Astronomy,
The Johns Hopkins University, Baltimore, MD 21218, USA}

\begin{abstract}

Chemical abundances studies of cataclysmic variables have revealed 
high nitrogen to carbon ratios in a number of of cataclysmic variable 
white dwarfs (based on ultraviolet emission and absorption lines), as well as 
possible carbon deficiency in many secondaries (based on the absence
of infrared CO absorption lines). These indicate that the accreted 
material on the white dwarf surface and the donor itself might be contaminated
with CNO processed material. To further understand the origin of
this abundance anomaly, there is a need for further chemical abundance
study. In the present work,  
we carry out a far ultraviolet spectral analysis of 
the extreme SU UMa dwarf nova TU Men and the U Gem dwarf nova SS Aur 
using archival spectra.  We derive the mass and temperature 
of the WD using the recently available DR2 Gaia parallaxes.      
The analysis of HST STIS spectra yields  
a WD mass  $M_{\rm wd}=0.77^{+0.16}_{-0.13}M_{\odot}$
with a temperature of $27,750 \pm 1000$~K for TU Men,  
and a WD mass $M_{\rm wd} \sim 0.80 \pm 0.15$ with a temperature
of $\sim 30,000 \pm 1000$~K for SS Aur. However, the analysis of a FUSE spectrum for SS Aur
gives to a higher temperature $\sim 33,375 \pm 1875 $~K, yielding to a higher   
WD mass $\sim 1 \pm 0.25 M_{\odot}$, which could be due to the effect of a 
second hot emitting component present in the short wavelengths of FUSE.  
Most importantly, 
based on the white dwarf far ultraviolet absorption lines,
we find that both systems have subsolar carbon and silicon 
abundances. For TU Men we also find suprasolar 
nitrogen abundance, evidence of CNO processing. 
 
\end{abstract}


\section{{\bf Introduction}}

The study of cataclysmic variables (CVs) has focused primarily on deriving
the system parameters (i.e. binary period, binary masses and temperatures, 
inclination, mass accretion rate,..) while less emphasis has been put 
on deriving the chemical abundances of the WD stellar photosphere.  
The main reason has been that to confront the theories of CVs evolution 
one primarily needs to know the binary period, 
the white dwarf (WD) mass, and the WD temperature
or the mass accretion rate for a large number of systems \citep[e.g.][]{pal17}.  
In addition, deriving stellar surface abundances demands {\it a priori}
high quality spectra and meticulous phase-resolved spectral analysis,
especially in the ultraviolet \citep[UV; e.g.][]{lon06,god17a}. 
In spite of that, chemical abundances studies of CVs have revealed 
high nitrogen to carbon ratios in a dozen CV WDs 
based on UV N\,{\sc v}/C{\sc iv} emission line ratios \citep[e.g.][]{gan03}  
and UV absorption lines \citep[e.g.][]{sio97,sio98,lon09,gan05}, as well as 
possible carbon deficiency in many secondaries \citep[based on the absence
of infrared CO absorption lines, see][for a short review]{how10}.  
From the study of a limited sample of CVs, it is estimated that possibly
up to 10-15\% of all CVs might have anomalously high N/C ratio 
\citep{gan03}.   
These results demonstrate the need for further studies to 
obtain stellar chemical abundances of the WD, the donor, and/or the disk.

In the present work, we present an analysis of 
the extreme SU UMa dwarf nova TU Men and the U Gem dwarf nova SS Aur  
as the first results of a   
much larger far-ultraviolet (FUV) analysis of CV systems  
to derive the chemical abundances in CV WDs.      
We chose these two systems as we previously analyzed them, they
have relatively good spectra revealing the white dwarf and 
exhibiting absorption lines, and,
as it appears {\it a posteriori}, they both show evidence of  
non-solar composition and possible CNO processed material. 

For that purpose, 
we analyze the medium quality/resolution short exposure time 
Hubble Space Telescope (HST) Space Telescope Imaging Spectrograph
(STIS) G140L/1425 snapshot spectra 
of TU Men and SS Aur.  
For SS Aur, we also analyze a Far Ultraviolet 
Spectroscopic Explorer (FUSE) spectrum. 
We improve on our previous analysis \citep{god08,sio08} of these two systems 
by considering a large range of values in the parameter space 
($Log(g),T_{\rm wd}$), using the Gaia DR2 derived distances,
and applying the reduced chi-squared statistic to find 
the best-fits.   
We further vary individual abundances of carbon, silicon, and nitrogen 
in the WD model atmosphere to assess the chemical abundances of 
the WD photospheres of these two systems. 
The STIS spectrum of TU Men reveals subsolar abundances of carbon 
and silicon as well as suprasolar abundance of nitrogen. 
The STIS spectrum of SS Aur too reveals subsolar abundances of
carbon and silicon, as to its FUSE spectrum it only show signs of 
subsolar abundance of carbon. 

We introduce the two systems in the next section and present the archival
data in section 3. The details of our spectral analysis tools
and technique are given in section 4. 
The results for the WD mass 
and temperature are presented in section 5 for both systems, 
while the results for the WD chemical abundances and projected stellar
rotational (broadening) velocity are given in section 6. 
We conclude with a discussion in the last section.

\begin{deluxetable}{lccc}[b!]  
\tablewidth{0pt}
\tablecaption{System Parameters
\label{syspar} 
} 
\tablehead{ 
Parameter   & Units         &     TU Men          &       SS Aur            
}
\startdata
Period      & (d)           &  0.1172$^{(1)}$     &  0.1828$^{(6,7)}$   \\[1pt] 
Inclination & (deg)         &  $44-70^{(1,5)}$    &  $38\pm16^{(8)}$     \\[1pt]
Distance    & (pc)          &  $278 \pm 5^{(2)}$  &  $260 \pm 3^{(2)}$   \\[1pt]
$E(B-V)$    &               &  $0.08\pm 0.02^{(4)}$  &  $0.08 \pm 0.02^{(3)}$        \\[1pt]   
WD Mass     & $(M_{\odot})$ &  ~~$0.6-1.06^{(1,5)}$ ~~&  ~~$1.08 \pm 0.40^{(8)}$  ~~ \\[1pt]  
$K_1$       & ~~(km/s) ~~   &  $85-150^{(1,5)}$   &  $70\pm10^{(6)}$       \\[1pt] 
\enddata
\tablecomments{ 
(1) \citet{men95};  
(2) derived from the Gaia DR2 parallax, \citet{ram17,lin18,lur18};  
(3) \citet{bru94}; 
(4) \citet{god17};
(5) \citet{sto84}; 
(6) \citet{sha86};
(7) \citet{kra65};   
(8) \citet{sha83}. 
} 
\end{deluxetable}

\section{{\bf The Two Systems}}  

\paragraph{{\bf TU Mensae.}} 
TU Men is a peculiar SU UMa system exhibiting some rather extreme
characteristics: it is the first SU UMa system discovered 
in the period gap with a period of $2^{h}.813 $ \citep{men95};
it has superoutburst maxima lasting for
more than 20 days or longer, occurring at extremely long intervals
\citep[$\sim years$,][]{bat00}; 
its disk is bright enough to induce
emission on the surface of the secondary star 
\citep[or alternatively it is the only SU UMa system to show emission 
from the secondary in quiescence,][]{tap03};   
and, with a mass ratio $\sim 0.5 \pm 0.2$, it belongs to a category of systems
exhibiting superhumps in spite of their longer orbital periods and higher mass
ratios  \citep[$q> q_{crit}=0.25$,][]{sma06}. 

Its WD mass is unknown and estimates varies from $\sim 0.6 M_{\odot}$
\citep{sto84}, $0.785 \pm 0.145 M_{\odot}$ to possibly  
$\approx 1.06 M_{\odot}$ \citep{men95}.  
Its $K_1$ amplitude was first measured from absorption lines in outburst
and found to be $\approx$150 km/s \citep{sto84}, while a measure from the 
emission lines at quiescence gave $K_1=108 \pm 10$ km/s or possibly 
$K_1=87 \pm 2$ km/s \citep{men95}.
These yielded, respectively, a mass ratio of q=0.59, 0.455, and 0.33, all
larger than the critical value of 0.25, below which the observed 
superhumps are believed to be due to the tidal instability of the 3:1 
resonance. It was shown \citep{sma06} that numerical simulations, on 
which the $q < 0.25$ criterion was based, do not apply to the observed
superhumps. 
Its inclination is possibly moderate and estimates have
put it at $65^{\circ} \pm 10^{\circ}$ \citep{sto84}, 
$52^{\circ} \pm 2^{\circ}$ or even as low as $\approx 44^{\circ}$
\citep{men95}. 
Well before Gaia, 
its distance was estimated to be 270 pc by fitting the observed
Balmer absorption lines to theoretical accretion disk lines 
for a mass transfer rate $\dot{M}=6 \times 10^{17}$g~s$^{-1}$,  
a white dwarf mass 
$M_{\rm wd} =0.6 M_{\odot}$, $q=0.588$, and $i=65^{\circ}$ 
\citep{sto84}. 
This is remarkably close to its Gaia DR2 derived distance of $278 \pm 5$ pc,
in spite of the fact the mass might be significantly larger, with a smaller
inclination and mass ratio. 
 
\paragraph{{\bf SS Aurigae.}} 
SS Aur is a {\it fairly normal} 
U Gem type dwarf nova with a period of $4^h.3872$ 
\citep{kra65,sha86}.  
The amplitude of the WD velocity, inferred from the variation
of the H$_{\alpha}$ emission lines, is $70 \pm 10$ km/s \citep{sha86}. 
The systems parameters were obtained by \citet{sha83}, but with a 
rather larger error for the WD mass ($1.08 \pm 0.40 M_{\odot}$) 
and inclination ($i=38 \pm 16 ^{\circ}$). 

A HST fine guidance sensors (FGS) parallax was obtained for SS Aur,
which, after calibration and correction, gave $4.97 \pm 0.65$ mas
\citep{har00}, or a distance of $201^{+30}_{-23}$ pc. 
However, the DR2 Gaia parallax of $3.849 \pm 0.041$ mas 
\citep{ram17,lin18,lur18} gives a distance of $260 \pm 3$ pc,
which we adopt here.    

While many non-magnetic CVs show possible carbon deficiency in 
their secondaries, based on the absence or weakness
of infrared CO absorption lines,
SS Aur is one of a few DNe presenting normal IR CO absorption 
lines \citep{how10}.  

The parameters of both systems are presented in Table \ref{syspar} 
with their source references.

\section{{\bf The Archival Data}} 

TU Men was observed with HST STIS on Jan 03, 2003, 
about 10 days after the end of an outburst. 
The STIS instrument was set up in the FUV configuration with the G140L grating
centered at 1425 \AA\ (with the 52"x0.2" aperture), 
thereby producing a spectrum from $\sim$1140 \AA\ to 
$\sim$1715 \AA\  (with a spectral resolution of $R\sim 1000$).  
The data (O6LI56010) were collected in ACCUM mode and consist in
one echelle spectrum only with a total good exposure time of 900 s.

SS Aur was observed with HST STIS on Mar 20, 2003,  
as part of the same ``snapshot survey'' as TU Men \citep{gan03,sio08},  
with the STIS instrument set up exactly in the same configuration.  
The data  (O6LI0F010)  were collected   
more than a year after the FUSE observations (see below), 
and 36 days after an outburst,  
totaling 600 s of good exposure time. 

All the HST STIS data were processed through the pipeline 
with CALSTIS version 2.13b. The STIS spectra of TU Men and SS Aur 
were presented in \citet{sio08} and \citet{god08}, respectively, 
with line identification.    

We also used recently retrieved data processed with CALSTIS version 3.4.2,
and extracted {\it new} spectra. 
We compare them to the data processed with CALSTIS version 2.13b 
\citep{god08,sio08} and found a very small change in the continuum
flux level reaching a maximum of 3\%.  
This is of the same amplitude as the systematic errors from
instrument calibration $\sim 3$\%.   
For comparison, the uncertainty in 
E(B-V) ($0.08 \pm 0.02$; see below) 
produces a change of 18\% in the continuum flux level
(see also Results Section 5).  
         
SS Aur was observed with FUSE on Feb 13, 2002 \citep{sio04}, at quiescence,
28 days after an outburst.   
The first dataset has an exposure time 
of $\sim$800 s, followed by 5 more data sets (obtained during successive 
FUSE orbits) with an exposure time nearing $\sim 2400$ s each,
totaling 12,733 s of good exposure time. 
The data were collected through the LWRS aperture in time tag (TTAG)
mode, and processed through the pipelines with the 
final version of CalFUSE \citep[v3.2.3;][]{dix07}.  
                   
The FUSE spectrum requires special attention, as the data come in the 
form of eight spectral segments which are combined together to give 
the final FUSE spectrum. The spectral segments were  individually 
examined \citep[][by visual inspection]{god08}  
to remove low-sensitivity portions, as well as other
instrument artifacts (such as the ``worm'') that cannot be corrected
by CalFUSE 
\citep[due to unpredicted temporal changes in strength and 
position of the artifact,][]{moo00,sah00,dix07}.  
Further details on the processing of the FUSE spectrum of SS Aur 
were given in \citet{god12}. 

\begin{deluxetable*}{lllcccrcc}[b!]  
\tablewidth{0pt}
\tablecaption{Archival Data Observation Log
\label{obslog}  
} 
\tablehead{ 
System & Instrument & Configuration & DATA ID   & Date (UT) & Time (UT) & Exp.time & Fig. &    Time since  \\ 
Name   &           &               &           & YYYY Mon DD & hh:mm:ss & sec     &  \#   &    outburst   
}
\startdata
TU Men & HST STIS/FUV & G140L (1425) & O6LI56010 & 2003 Jan 04 & 01:47:53 &  900  &  \ref{tumenfit},\ref{tumen_silicon},\ref{tumen_carbon},\ref{tumenab}  & 10 d  \\  
SS Aur & HST STIS/FUV & G140L (1425) & O6LI0F010 & 2003 Mar 20 & 11:39:59 &  600  &  \ref{ssaurstis},\ref{ssaurstisab}  &    36 d  \\ 
       & FUSE         & LWRS         & C11002010 & 2002 Feb 13 & 06:49:53 & 12733 &  \ref{ssaurfuse},\ref{ssaurfusedetail}   &  28 d  \\  
\enddata
\tablecomments{
The date/time refer to the start of the observation,
the exposure time is the total good exposure time.  
}   
\end{deluxetable*}

The FUSE spectra are subject to larger systematic errors than
the STIS spectra, with an uncertainty in the FUSE flux calibration
estimated at $\sim$10-15\% (FUSE data handbook v1.1). 
More specifically, data that were obtained early in the mission
(as is the case for SS Aur) might suffer from a measured flux
that is too high by 5-10\% in the LiF2A channel (1095-1135 \AA), 
and the LiF1B channel (covering about the same wavelength as the 
LiF2A) can be off by up to 10\% (FUSE DATA Handbook 2009).  
An explicit comparison of the same FUV sources \citep{boh14} shows that 
FUSE spectra exhibit a local flux variation of $\pm 5$\% 
compared to STIS spectra. We can, therefore, assume a {\it systematic
error} of at least $\sim 10\%$ in the continuum flux level 
for FUSE, and possibly as high as $\sim 15\%$.

With a wavelength coverage from $\sim 905$ \AA\ to $\sim 1095$ \AA , 
the FUSE telescope was the ideal instruments to study the hot 
accreting WDs in CVs.  
However, FUSE spectra are often contaminated with 
interstellar medium (ISM) absorption lines dominated by atomic 
and molecular hydrogen lines \citep[e.g.][]{sem01},
as the Lyman series extends from $\sim 912$ \AA\ to 1216 \AA\ .  
Because of its location near the galactic plane,   
and in spite of it being only at 260 pc, 
SS Aur presents some prominent ISM 
absorption lines in its FUSE spectrum.  
Explicitly, these lines consist in the W-Wener and L-Lyman band, 
upper vibrational level (1-16), and 
rotational transition (R,P, or Q with lower rotational state 
J=1-3). Additional ISM lines include lines from  
N\,{\sc i} \& N\,{\sc ii}, Fe\,{\sc ii}, 
Si\,{\sc ii}, and C\,{\sc ii}.
The ISM absorption lines in the FUSE spectrum of SS Aur
were identified in \citet{god08}.  

The observation log of both the FUSE and HST archival data 
is presented in Table \ref{obslog}.

In preparation for the fitting, we deredden the spectra  
assuming E(B-V)=0.08 (see Table \ref{syspar}). 
For the extinction curve, 
we use the analytical expression of \citet{fit07}, that we have 
slightly modified to agree with an extrapolation of \citet{sav79}
in the FUSE range.  \citet{sas02} have shown \citep[see also][]{sel13}
that in the FUV range the observed extinction 
curve is actually consistent with an extrapolation of the standard extinction 
curve of \citet{sav79}.  
For many CVs the reddening value has been estimated by ``ironing out'' the 
2175~\AA\ extinction bump \citep[e.g.][using {\it International Ultraviolet
Explorer} (IUE) spectra]{ver87}. However, the normalized height of the 
2175~\AA\ feature has an accuracy of about $\pm 20$\% and it
has been suggested \citep{fit99} that, as a consequence, the uncertainty in 
$E(B-V)$ (derived from ironing out the bump) must have a similar relative
uncertainty. In the present case we assume an uncertainty of 
$\pm 0.02$ for a reddening $E(B-V)=0.08$ (the two systems have the
same value), or $\pm 25$\%.

\section{{\bf FUV Spectral Analysis Tools and Technique}} 

We use the suite of codes TLUSTY/SYNSPEC \citep{hub88,hub95} 
to generate synthetic spectra for high-gravity stellar atmosphere 
WD models. A one-dimensional vertical stellar atmosphere structure is 
first generated with TLUSTY for a given surface gravity ($Log(g)$),
effective surface temperature ($T_{\rm wd}$), and surface composition. 
Subsequently, the code SYNSPEC is run, using the output from TLUSTY 
as an input, to solve for the radiation field and generate a synthetic 
stellar spectrum over a given wavelength range between 900~\AA\ and 7500~\AA . 
The code includes the treatment of the quasi-molecular satellite lines
of hydrogen which are often observed as a depression around 1400~\AA\ 
and around 1060~\AA\ and 1080~\AA\ 
in the STIS and FUSE spectra (respectively) 
of WDs at low temperatures and high gravity: 
$T_{\rm wd} \le 30,000$ K and $Log(g) \ge 7.50$, 
\citep[e.g. see][]{god08}.
For temperatures above 35,000~K, we switch on the NLTE option with lines, 
considering all lines of H and He explicitly. 
Last, the code ROTIN is used to reproduce rotational and instrumental
broadening as well as limb darkening. 
Technical and practical details on the suite of codes TLUSTY/SYNSPEC
are given in \citet{hub17a,hub17b,hub17c}, and the FORTRAN programs
(with input files and examples) are downloadable from the TLUSTY
webpage\footnote{\url{http://tlusty.oca.eu/}}.          

In this manner we have already generated solar composition  
stellar photospheric spectra covering a wide range of effective temperatures
$T_{\rm wd}$ and surface gravities $Log(g)$:  
from $\sim$18,000~K to $\sim$40,000~K in steps of 1000~K, 
and an effective surface gravity from $Log(g)=7.0$ to $Log(g)=9.0$ 
in steps of 0.2.  
For each WD temperature and gravity there is a single WD radius $R_{\rm wd}$  
and mass $M_{\rm wd}$, obtained using the non-zero temperature C-O WD 
mass-radius relation from \citet{woo95}.

The fitting of the observed spectra with theoretical spectra is carried
out in two distinct steps: (i) in the first step we  derive the WD 
effective surface temperature and gravity by fitting the continuum 
and the hydrogen Ly$\alpha$ profile; 
(ii) in the second step we derive the stellar surface abundances and 
the projected stellar rotational velocity by fitting the 
metal absorption lines.  

\subsection{\bf{Fitting the Continuum and Hydrogen Ly$\alpha$ Profile.}} 

Using our existing theoretical model spectra with solar composition
and a {\it standard} projected rotational velocity of $\sim$200 km/s for
TU Men (which is common for cataclysmic variables) and $\sim$400 km/s 
for SS Aur \citep{god08,sio08}, we fit the observed spectra using the  
$\chi^2_{\nu}$ minimization technique to find the best fit. 
Since we fit the continuum and the Lyman orders profile, the exact
value of the velocity broadening (here the projected rotational velocity) 
does not affect the results. The correct value of the velocity is found
in the second step together with the chemical abundance of the elements. 
The distance $d$ is also obtained for each model 
by scaling the model to the observed spectrum. 
Doing so we obtain the reduced $\chi^2_{\nu}$ 
($\chi^2$ per degree of freedom $\nu$) and $d$ in the   
$T_{\rm wd}$ vs. $Log(g)$ parameter space. 
A preliminary check shows that for both TU Men and SS Aur, 
the least $\chi^2$ solutions
for the known Gaia distance are obtained for $Log(g) > 7.5$, and 
for a temperature  of 25,000-35,000~K. We also find that 
the observed spectra are better fitted without the inclusion of the
hydrogen quasi-molecular satellite line. 

We, therefore, refine the grid of models and generate additional 
theoretical spectra, without the inclusion of the hydrogen quasi-molecular
satellite lines, in the parameter space $(Log(g),T_{\rm wd})$: 
for $ 7.4 \le Log(g) \le 9.0$, in steps of  0.1 in $Log(g)$,  
and  for 23,000 K $ \le  T_{\rm wd} \le$ 37,000 K,
in steps of 500~K in $T_{\rm wd}$.  
The refined grid consists of a total of 
493 ($17 \times 29$) {\it solar composition} 
theoretical WD spectra.           
We then carry out a new spectral to fit the observed spectra
to obtain the least $\chi^2_{\nu}$ solutions in the two-dimensional
parameter space $T_{\rm wd}$ vs $Log(g)$.

\subsection{\bf{ 
The $\chi^2$ Minimization Procedure and The Statistical Errors 
}} 

We use a standard least-squares minimization approach to find the best-fit 
model and derive $T_{\rm wd}$ and $Log(g)$.  
In theory a good fit is obtained for a (reduced) chi-square $\chi^2_{\nu} \sim 1$,
though in practice the least $\chi^2_{\nu~MIN}$ value can be smaller or larger than one. 
The value of $\chi^2$ is also subject to noise, which is inherited from the noise of the data. 
As a consequence there is an uncertainty in $\chi^2$ (and therefore $\chi^2_{\rm MIN}$), which translates into 
uncertainties on the derived parameters $Log(g)$ and $T_{\rm wd}$ - the {\it statistical errors}.  
For a number $p$ of parameters (here $p=2$, for $T_{\rm wd}$ and $Log(g)$), 
the uncertainty on the parameters $p$ is obtained for $\chi^2$ within the range 
$\chi^2_{\rm MIN} + \chi^2_p (\alpha)$,
where $\chi^2_p(\alpha)$ is given in Table 3 for a significance $\alpha$,  
or equivalently, for a confidence $C=1-\alpha$ 
\citep[see e.g.][]{lam76,avn76}. 

\begin{deluxetable}{ccccc}[h!]  
\tablewidth{0pt}
\tablecaption{Uncertainty in $\chi^2$ with $p$ Parameters and Significance $\alpha$ with $\nu$ 
Degrees of Freedom  
\label{chierrors} 
} 
\tablehead{ 
\multicolumn{1}{c}{Significance} & \multicolumn{1}{c}{Confidence} & \multicolumn{3}{c}{$\chi^2_p(\alpha)$}   \\\cmidrule{3-5}  
\multicolumn{1}{c}{$\alpha$}     & \multicolumn{1}{c}{C}          & \multicolumn{1}{c}{$p=1$} & \colhead{$p=2$}  & \multicolumn{1}{c}{$p=3$}  
}
\startdata
            0.32       &  0.68~(1.0$\sigma$)      &   1.00       &   2.30   &  3.50  \\       
            0.10       &  0.90~(1.6$\sigma$)      &   2.71       &   4.61   &  6.25  \\       
            0.01       &  0.99~(2.6$\sigma$)      &   6.63       &   9.21   &  11.3  \\       
\enddata
\tablecomments{If the data consist of $N$ data points to be fitted, 
and there are $p$ independent parameters,  one has $ \nu = N -p$. 
The uncertainty in the reduced chi square $\chi^2_{\nu}$ is obtained 
by dividing $\chi^2_p(\alpha)$ by the number of degrees of freedom $\nu$.  
} 
\end{deluxetable}

In the present work we derive the statistical errors on the best fit
$T_{\rm wd}$ and $Log(g)$ for a 99\% (2.6 $\sigma$) confidence level. 
The $p=2$ case  
is to be used when finding the error on the best-fit model in the 
two-dimensional parameter space $(T_{\rm wd},Log(g))$. 
However, as soon as one uses the known distance $d$ as a constraint, 
the problem is reduced to a one-parameter problem: 
one has to find the best fit along the line $distance=d$ in the 
$(T_{\rm wd},Log(g))$  parameter space. In that case one uses 
the $p=1$ case.  
We will come back to this in the result section.

\subsection{\bf{Fitting the Metal Absorption Lines.}} 

In the second step, after we found the best   
$Log(g)$ and $T_{\rm wd}$ fit for each spectrum, 
we vary the abundances of the elements 
Si, S, C, \& N, one at a time, and vary the projected stellar rotational 
velocity, $V_{\rm rot} sin(i)$ in the best fit model. 
The abundances are varied from $0.01 \times$ solar to 
$50 \times$ solar in steps of about a factor of two or so; 
the projected stellar rotational velocity is varied  from
50 km/s to 1000 km/ in steps of 50 km/s. 
The results of the abundances/velocity are then examined by visual 
inspection of the fitting of the absorption lines for each element. 
The reason we use visual examination rather than the $\chi^2$ minimization
technique is 
that a visual examination can recognize and distinguish 
real absorption features from the noise, as the data quality is modest. 
This is partly because the spectral
binning size ($\sim 0.58$~\AA\ for STIS) is of the same order of magnitude  
as the width of some of the absorption lines, 
and the depth of some of the absorption features is of the 
amplitude as the flux errors. 
This is explicitly shown in Figs.  \ref{tumen_silicon} \& \ref{tumensierrors} 
in the Results section 6.
Details on our technique to derive stellar abundances and broadening
velocities are also given in \citet{god17a}. 

\clearpage 

\section{{\bf The White Dwarf Surface Temperature and Gravity.}}

\subsection{{\bf TU Men}} 

\begin{figure}[h!]
\vspace{-19.cm} 
\plotone{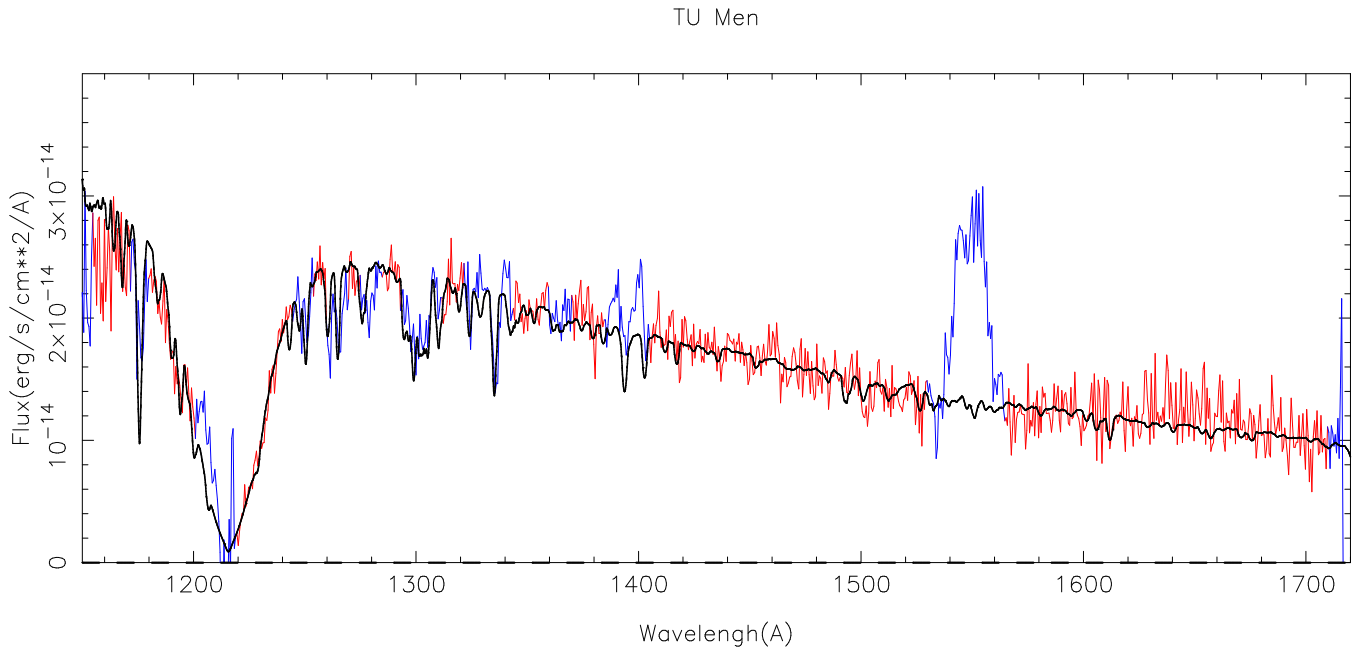} 
\vspace{-4.cm} 
\caption{
Modeling of the WD in TU Men.
The HST STIS FUV spectrum of TU Men (in red) is fitted with 
a synthetic stellar spectrum (in black). 
The STIS spectrum has 
been dereddened assuming $E(B-V)=0.08$. A first fit is carried out 
where the  emission lines and absorption lines are masked to find
the temperature and gravity, these regions are in blue. 
Once the best-fit temperature and
gravity are found, the absorption lines are fitted to find the 
abundances and projected rotational velocity.  
The WD model has an effective 
surface gravity of $Log(g)=8.25 \pm 0.25$, with an effective
surface temperature of $27,750 \pm 1000$~K. 
The model has a carbon abundance $[C] = 0.2 \pm 0.1$, silicon abundance
$[Si]=0.2 \pm 0.1$, and nitrogen abundance $[N]=20 \pm 10$,
all in solar abundance units (sun=1). 
The projected stellar rotational velocity of the model 
is $V_{rot} \sin(i) = 225 \pm 75 $ km/s. 
\label{tumenfit} 
}
\end{figure}

We first present the results for the WD  
{\it temperature} and {\it gravity} obtained from the spectral
fit to the STIS spectrum of TU Men dereddened assuming $E(B-V)-=0.08$. 
Before the fitting we mask all the emission lines
(including the Si\,{\sc iv} doublet), the bottom of the Ly$\alpha$, 
as well as the absorption lines (such as e.g. the C\,{\sc iii} 1175 \AA , 
Si $\sim$ 1300 \AA\ feature) since these depends on the 
exact abundances. We also mask a portion of the left wing of the 
Ly$\alpha$ as this region 
presents a possible emission region (near 1200 \AA ).  
We must stress that the masking of these spectral regions helps decrease 
the minimum $\chi^2_{\nu}$ (best fit) to a value close to 1. 
The best-fit model (Fig.\ref{tumenfit}, with the masked regions in blue)
has an effective surface gravity of $Log(g)=8.25 \pm 0.25$, with an effective
surface temperature of $27,750 \pm 1000$~K. 
The model has a carbon abundance $[C] = 0.2 \pm 0.1$, silicon abundance
$[Si]=0.2 \pm 0.1$, and nitrogen abundance $[N]=20 \pm 10$,
all in solar abundance units (sun=1). 
The projected stellar rotational velocity of the model 
is $V_{rot} \sin(i) = 225 \pm 75 $ km/s. 

In the next paragraphs give a full account 
of how the best-fit model was obtained together with estimates of the 
systematic and statistical errors.

\clearpage

\begin{figure}[b!]
\vspace{-5.cm} 
\plottwo{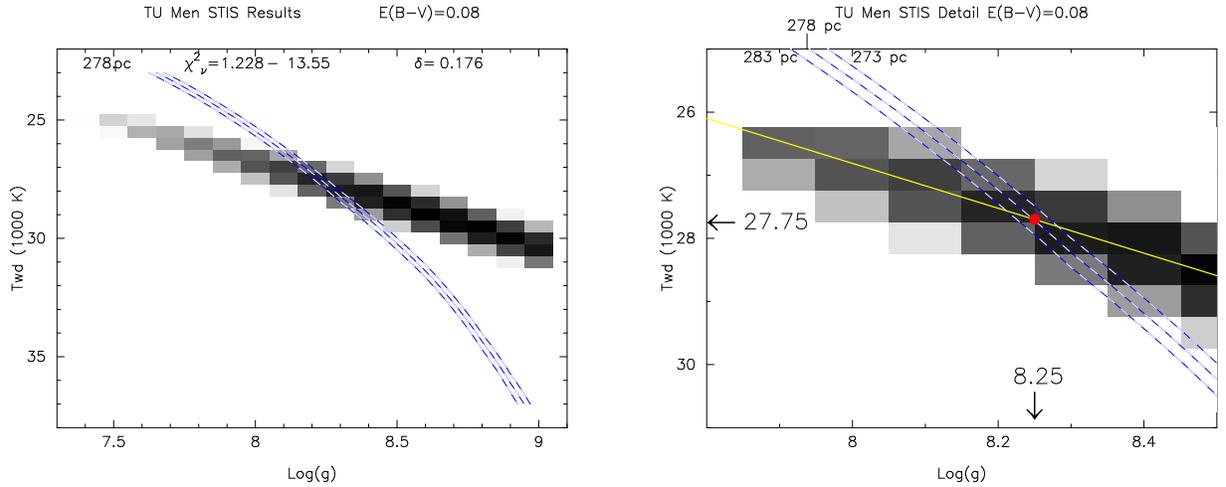}{tumen_stis_chi_ebv0.08_detail.eps}  
\caption{ 
{\bf Left.} The results of the spectral analysis of 
the STIS spectrum of TU Men are displayed
in the {\it decreasing} $T_{\rm wd}$ vs. {\it increasing} $Log(g)$ 
parameter space as a map of the $\chi_{\nu}^2$ value. 
The size of each model is that of a (gray) rectangle of 500 K by 0.1 in Log(g).
The least $\chi^2$ models are within the gray diagonal area. 
The other models in white have fast increasing $\chi^2_{\nu}$ values 
towards the lower left and upper right,  
reaching a maximum value of 13.55.
The blue-white dashed triple line indicates a 
scaling to the Gaia distance of $278 \pm 5$ pc: 
273 pc (right dashed line), 278 pc (center dashed line), and 283 pc 
(left dashed line).  
\\  
{\bf Right.} A detailed (zoom-in) of the left panel shows 
the intersection of the least $\chi_{\nu}^2$ models 
with the Gaia distance estimates. The distance has to be between 
273 pc (right blue dashed line) and 283 pc (left blue dashed line).
The smallest values of $\chi^2_{\nu}$ are obtained on the yellow
diagonal. 
The resulting intersection of the 278~pc distance and the least
$\chi^2_{\nu}$ gives the best fit (represented with a red dot)  
$Log(g)=8.25$, with $T_{\rm wd}=27,750$~K.
\label{tumenchi} 
}
\end{figure}

The overall results are summarized in Fig.\ref{tumenchi}, which were obtained 
using the refined grid of models (Sec.4). 
On the left panel we draw a grayscale 
of the $\chi^2_{\nu}$ value in the parameter space $(Log(g),T_{\rm wd})$.  
The least $\chi^2_{\nu}$ values (best fit models, in different shades of gray) 
form a diagonal starting at
25,000 K (for $Log(g)=7.5$) and extending to 30,500 K (for $Log(g) = 9.0$).  
These models have 
$ \chi^2_{\nu {\rm MIN}} \le \chi^2_{\nu} \le \chi^2_{\nu {\rm MIN}} + \delta $, 
where $\chi^2_{\nu {\rm MIN}} = 1.228$ and $\delta = 0.176$. 
The other models have fast increasing $\chi^2_{\nu}$ values 
when moving away from that diagonal (they have been left in white), 
reaching a maximum $\chi^2_{\nu}=13.55$.
The area of interest is where the best fit models (gray diagonal) scale to the
Gaia distance. 
An enlarged view of that area is shown on the right panel of Fig.\ref{tumenchi}, 
where we draw a yellow line at the center of the gray diagonal. 
The intersection of the yellow line with $d=278$~pc gives the solution 
$Log(g)=8.25$ with $T_{\rm wd}=27,750$~K. 
Due to the finite size of the grid of models,  
$500$~K in $T_{\rm wd}$ and $\pm 0.10$ in $Log(g)$, 
there is a intrinsic uncertainty in the results reaching a maximum  
of $\pm 250$~K in $T_{\rm wd}$ and $\pm 0.05$ in $Log(g)$. 

Next, we estimate how the uncertainty in the Gaia distance
($\pm 5$~pc) affects the results by checking where the 
yellow line (still in Fig.\ref{tumenchi}, right panel) intersects 
the blue-white dashed lines representing the distances $d=283$~pc 
and $d=273$~pc. We obtain an uncertainty of $\pm 100$~K in $T_{\rm wd}$
and $\pm 0.02$ in $Log(g)$.

\clearpage 

\begin{figure}[b!]
\plottwo{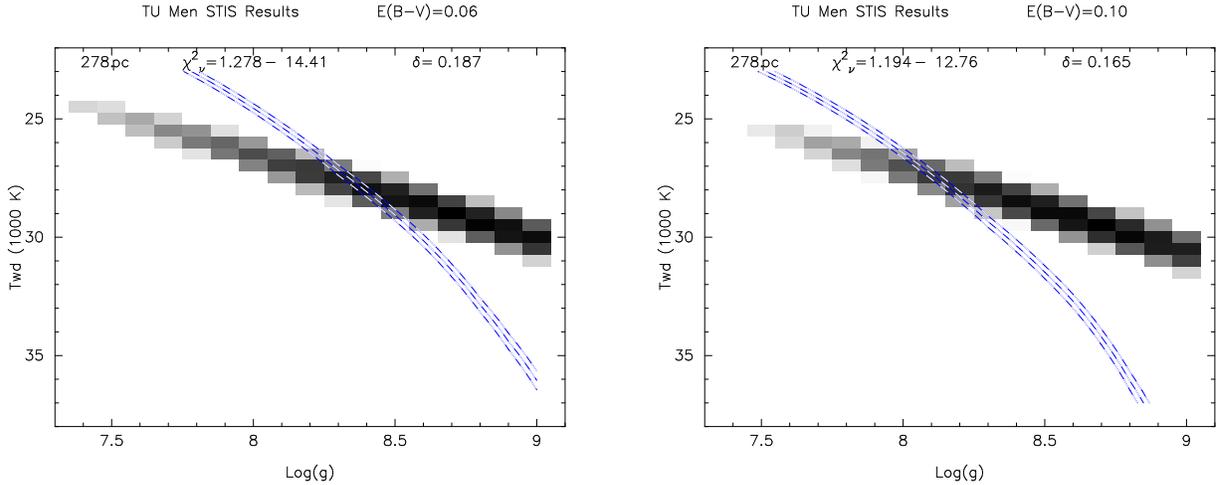}{tumen_stis_chi_ebv0.10.eps}  
\caption{ 
Map of $\chi^2_{\nu}$ for the fitting of the STIS spectrum of TU Men
assuming $E(B-V)=0.06$ (left panel) and $E(B-V)=0.10$ (right panel). 
The gray scale, and dashed lines have the same meaning
as in Fig.\ref{tumenchi}. 
The best fit models (gray diagonal) have a slightly lower 
temperature ($\approx -250$~K) 
for $E(B-V)=0.06$ and a slightly higher temperature ($\approx +250$~K) 
for $E(B-V)=0.10$ than for $E(B-V)=0.08$. 
Due to the scaling of the flux to the distance, the triple dashed 
line of the Gaia distance also shifts due to the change in the assumed 
value of $E(B-V)$. As a consequence 
the best fit is $T_{\rm wd}=28,000$~K and $Log(g)=8.4$ for $E(B-V)=0.06$
(on the left),  
and $T_{\rm wd}=27,500$~K and $Log(g)=8.1$ for $E(B-V)=0.10$ (on the right). 
Namely, the uncertainty of $\pm $25\% in $E(B-V)$ gives an uncertainty
of $\pm 250$~K in $T_{\rm wd}$ and $\pm 0.15$ in $Log(g)$.   
\label{tumenebminusv} 
}
\end{figure}

To assess how the uncertainty in the reddening value/extinction law (Sec.3)   
propagates, we carry out the same analysis with the STIS spectrum of TU Men 
dereddened assuming now $E(B-V)=0.06$ and $E(B-V)=0.10$. We find, Fig.\ref{tumenebminusv}, 
that the uncertainty of $\pm 0.02$ in $E(B-V)$ produces an uncertainty
of $\pm 250$~K in $T_{\rm wd}$ and $\pm 0.15$ in $Log(g)$.    
The rather large uncertainty produced in $Log(g)$ is due to the relative
shift between the (dashed) line of the distance and the least $\chi^2$ gray diagonal 
in {\it opposite direction} along the $Log(g)$ axis. 
The moderate uncertainty of $\pm 250$~K is due to their relative shift  
in the {\it same direction} along the $T_{\rm wd}$ axis. 
    
We recall that the errors from the STIS instrument/calibration are only 3\% in continuum 
flux level (see Sec.3), the same order of magnitude as the errors from the Gaia distance.
These errors are negligible compared 
to the uncertainty in flux (18\%) propagating from the reddening uncertainty.

\clearpage

\begin{figure}[h!]
\vspace{-10.cm} 
\plotone{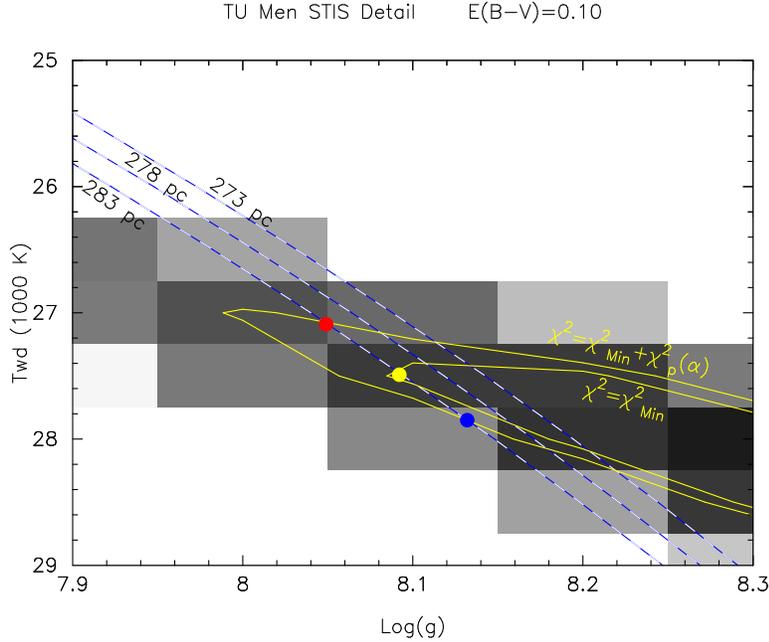} 
\vspace{3.5cm} 
\caption{Map of $\chi^2$ with statistical errors. 
In order to assess the statistical errors we consider the least
$\chi^2_{\rm Min}$ solution for the $d=283$~pc distance, displayed here
with the yellow dot. We draw the contour for that value of 
$\chi^2 = \chi^2_{\rm Min}$. 
The statistical errors is then obtained by considering the region for which 
$\chi^2_{\rm Min} \le \chi^2 \le \chi^2_{\rm Min} + \chi^2_p(\alpha)$ for $p=1$
and $\alpha=0.01$ (see Table 3). The errors on $T_{\rm wd}$ and $Log(g)$
are then obtained by considering the intersections of the 
$\chi^2_{\rm Min}+\chi^2_p(\alpha)$ with the $d=283$~pc (dashed) line
(the red and blue dots). 
Overall, the statistical errors have a magnitude
of $\pm 350$~K in $T_{\rm wd}$ and $\pm 0.04$ in $Log(g)$.
\label{statistical}
} 
\end{figure}

We now turn to the statistical errors, from the $\chi^2$ minimization
technique,  for which we consider the 
problem to be one-dimensional. Namely, while the solution are presented
in the two-dimensional parameter space $T_{\rm wd}$ vs $Log(g)$, 
the constraint $d=278$~pc transforms the problem
into a one-dimensional problem: one has to find the least $\chi^2$
value along the white-blue dashed line representing the distance.    
Once one finds the minimum chi square value $\chi^2_{\rm MIN}$ 
along the distance line, the statistical uncertainties on the parameters 
$T_{\rm wd}$ and $Log(g)$ are obtained by considering all the solutions 
for which 
$\chi^2_{\rm Min} \le \chi^2 \le \chi^2_{\rm Min} + \chi^2_p(\alpha)$, 
with $p=1$ and $\alpha=0.01$ (99\% confidence) as in Table 3 (see Section 4.2). 
In Fig.\ref{statistical} we display such a map of $\chi^2$ with 
the statistical uncertainty for the distance line $d=283$~pc 
(the choice of this distance will become clear in the next paragraph).  
We obtain a statistical uncertainty of  $\pm 350$~K in $T_{\rm wd}$, 
and $\pm 0.04$ in $Log(g)$. 
We recapitulate all the errors in Table \ref{errors}. 

The way we assess the cumulative effect of the different uncertainties 
is illustrated in Fig.\ref{statistical}. 
The red dot represents the lowest temperature and lowest gravity
of the solution due to propagation of the cumulative uncertainties of the 
(i) distance, (ii) reddening and (iii) statistical errors.  
Similarly, the $\chi^2_{\rm Min}+\chi^2_p(\alpha)$, $d=273$~pc 
for $E(B-V)=0.06$ gives the highest temperature and highest gravity
within the margin of errors.    
Including the known quantifiable errors in $E(B-V)$, $d$, and $\chi^2$,
we obtain $T_{\rm wd} = 27,750 \pm 700$~K, and $Log(g)=8.25 \pm 0.20$.

\clearpage 

If we also include the errors introduced from the STIS/instrument 
and our own modeling we have:  
$T_{\rm wd} = 27,750 \pm 1000$~K, and $Log(g)=8.25 \pm 0.25$, 
where we have slightly rounded down the errors.
All the errors are listed in Table \ref{errors}. 

Using the mass-radius relation for a $\sim 28,000$K WD \citep{woo95}, 
we find that $Log(g)=8.25 \pm 0.25$ corresponds to a WD mass 
$M_{\rm wd} \approx 0.77^{+0.16}_{-0.13} M_{\odot}$.

\begin{deluxetable}{lrccccc}[h!]  
\tablewidth{0pt}
\tablecaption{Results for $T_{\rm wd}$ and $Log(g)$ with Errors Estimates  
\label{errors} 
} 
\tablehead{ 
\multicolumn{1}{c|}{         } & \multicolumn{2}{c|}{TU Men} & \multicolumn{2}{c|}{SS Aur} & \multicolumn{2}{c}{SS Aur} \\ 
\multicolumn{1}{c|}{         } & \multicolumn{2}{c|}{STIS} & \multicolumn{2}{c|}{STIS} & \multicolumn{2}{c}{FUSE} \\ 
\multicolumn{1}{c|}{ } & \multicolumn{1}{c}{$T_{\rm wd}$} & \multicolumn{1}{c|}{$Log(g)$} & \colhead{$T_{\rm wd}$} & \multicolumn{1}{c|}{$Log(g)$} & \colhead{$T_{\rm wd}$} & \multicolumn{1}{c}{$Log(g)$} \\\cmidrule{1-7}   
\multicolumn{1}{c|}{Best-fit}  & 27,750    & 8.25       & 30,000    & 8.275      & 33,375    &  8.66      \\ \cmidrule{1-7} 
\multicolumn{1}{c}{}                           & \multicolumn{6}{c}{ Errors  }   \\\cmidrule{1-7}  
\multicolumn{1}{c|}{Source of Errors} & \multicolumn{1}{c}{$\Delta T_{\rm wd}$} & \multicolumn{1}{c|}{$\Delta Log(g)$} & \colhead{$\Delta T_{\rm wd}$} & \multicolumn{1}{c|}{$\Delta Log(g)$} & \colhead{$\Delta T_{\rm wd}$} & \multicolumn{1}{c}{$\Delta Log(g)$}    
}
\startdata
distance $d$ & $\pm 100$ & $\pm 0.02$ & $\pm 100$ & $\pm 0.02$ & $\pm 100$ & $\pm 0.02$ \\  
$E(B-V)$     & $\pm 250$ & $\pm 0.15$ & $\pm 250$ & $\pm 0.15$ & $\pm 250$ & $\pm 0.15$ \\  
instrument   & $\pm 100$ & $\pm 0.02$ & $\pm 100$ & $\pm 0.02$ & $\pm 500$ & $\pm 0.15$ \\  
modeling     & $\pm 250$ & $\pm 0.05$ & $\pm 250$ & $\pm 0.05$ & $\pm 250$ & $\pm 0.05$ \\  
statistical $\chi^2$ & $\pm 350$ & $\pm 0.04$ & $\pm 250$ & $\pm 0.025$& $\pm 775$ & $\pm 0.06$ \\  
\enddata
\tablecomments{The temperature is in Kelvin and the gravity in cgs. 
} 
\end{deluxetable}

\clearpage 

\subsection{{\bf SS Aur.}} 

\subsubsection{{\bf The STIS Spectral Analysis.}} 

\begin{figure}[h!]
\vspace{-19.cm} 
\plotone{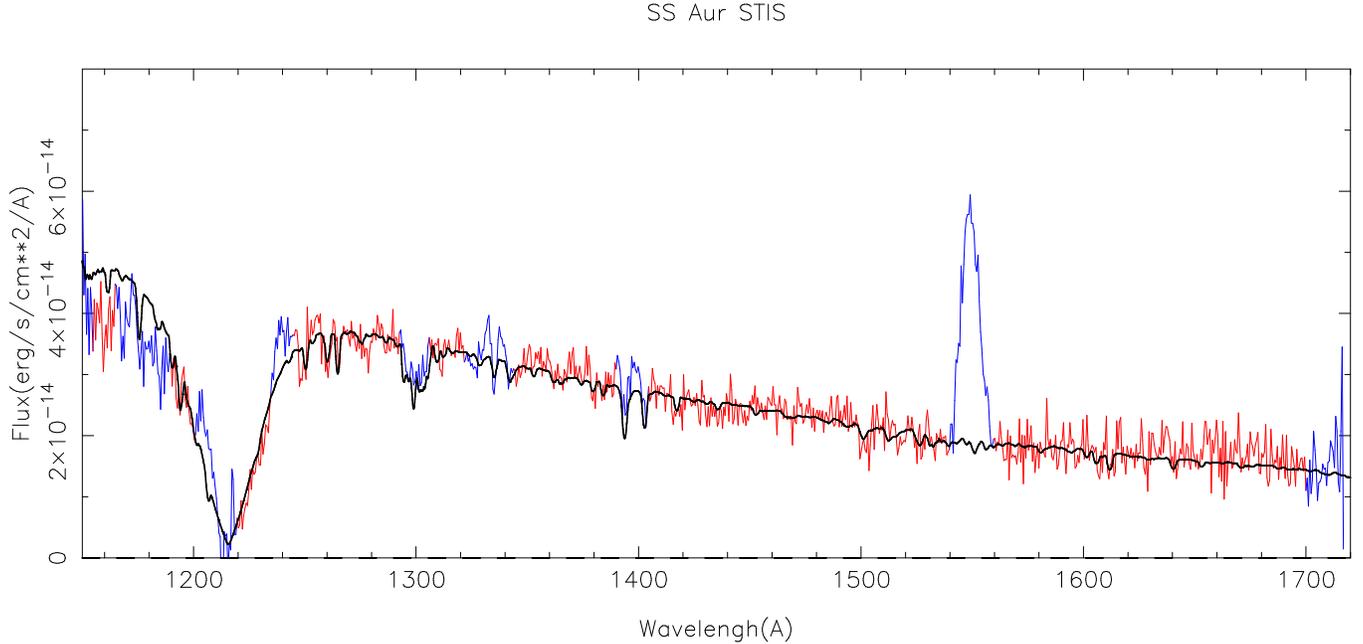} 
\vspace{-4.cm} 
\caption{ 
A WD best-fit model (black line) 
to the STIS spectrum of SS Aur (in red) obtained
at quiescence on 2003 Mar 20, 36 days after an outburst. 
The STIS spectrum exhibits a strong C\,{\sc iv} ($\sim$1550) emission line, 
and possibly some weak N\,{\sc v} ($\sim$1240), C\,{\sc ii} ($\sim$1336), and
Si\,{\sc iv} ($\sim 1400$) emission features. 
Before the WD temperature and gravity fit, we mask emission and absorption
lines (in blue). We also mask most of the region below 1200~\AA\ , as 
the observed flux is significantly smaller than the models and could
not be fitted.  
The model has $T_{\rm wd}=30,000$~K 
for $Log(g)=8.275 $ ($M_{\rm wd} = 0.785 M_{\odot}$). 
Fitting the absorption lines is given in section 6, though this model 
has a projected stellar rotational velocity $V_{\rm rot} sin(i) = 400$ km/s,   
a silicon abundance [Si]=0.2 solar, and a very low
carbon abundance (0.001 solar). 
\label{ssaurstis}    
}
\end{figure}

For the STIS spectrum of SS Aur, we mask the C\,{\sc ii} 
(1336) and C\,{\sc iv} (1550) emission lines, as well as the Si\,{\sc iv} 
(1400) doublet emission feature. We also mask the prominent absorption lines
and a portion of the left wing of the Ly$\alpha$ region due to an apparent
emission line there.  There is no evidence of the 
C\,{\sc iii} (1175) absorption line, we therefore also mask that region. 
The masked regions are shown in Fig.\ref{ssaurstis} for the best-fit model, 
for which $Log(g)=8.275$ and $T_{\rm wd} = 30,000$ K.

\clearpage 

\begin{figure}[t!]
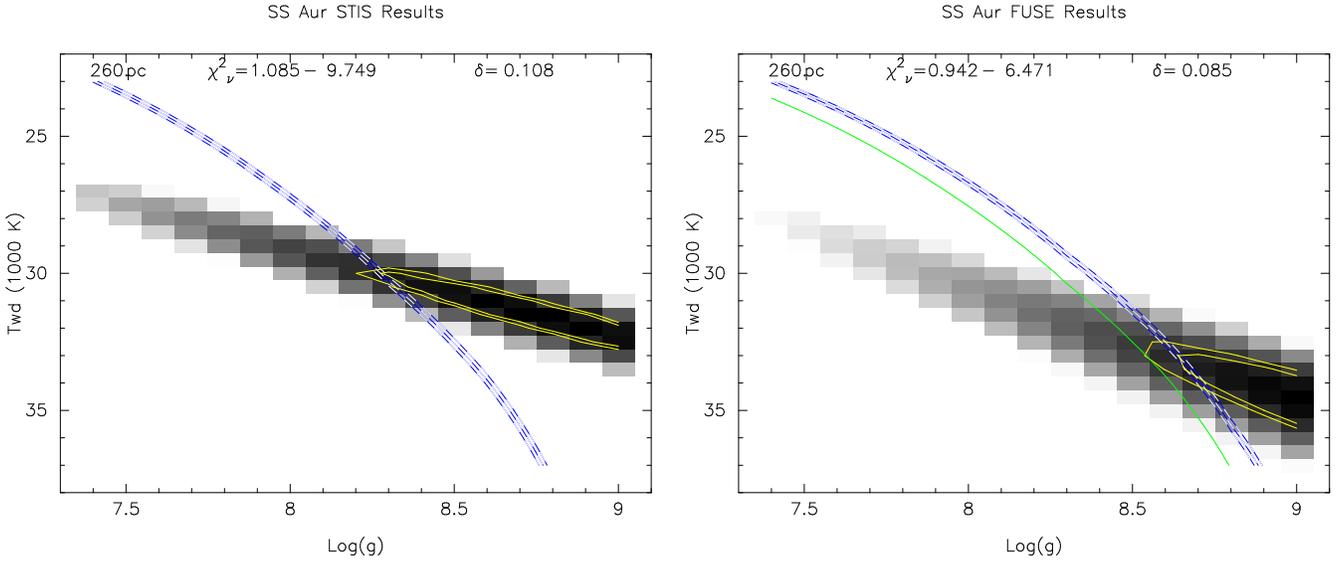

\vspace{-2.cm}
\gridline{\fig{ssaur_stis_chi.eps}{0.49\textwidth}{}
          \fig{ssaur_fuse_chi.eps}{0.49\textwidth}{}
}
\vspace{-1.cm}
\caption{
The results of the spectral analysis of 
the STIS (left panel) and FUSE (right panel) spectra of SS Aur are displayed
in the $(log(g),T_{\rm wd})$ parameter space as a map of the 
$\chi_{\nu}^2$ value, as in Fig.\ref{tumenchi}. 
The yellow contour lines 
are for the assessment of the statistical errors as in Fig.\ref{statistical}.  
Because the FUSE and STIS
spectra have slightly different flux levels and different wavelength
coverage, the distance (triple blue dashed lines) 
does not cross the parameter space in the same manner in both panels.     
For the STIS spectrum we obtain (left)  
$Log(g)=8.275$ with $T_{\rm wd} = 30,000$ K. 
For the FUSE spectrum (right) 
the least-$\chi_{\nu}^2$ model with the correct distance yield 
$Log(g)=8.66$ with $T_{\rm wd} = 33,375$~K. 
The green line in the right panel 
represents the solution if the continuum flux level 
in the FUSE spectrum was lowered by 15\%, the maximum possible 
(though unlikely) error due to the FUSE instrument. 
For a full treatment of the errors in both the FUSE and STIS
analyses, see text.
\label{ssaur_chi} 
}
\end{figure}

We follow the same procedure as for TU Men to assess the best-fit
and the errors.  The overall results for the STIS spectrum of SS Aur 
are presented in Fig.\ref{ssaur_chi} for $E(B-V)=0.08$, left panel. 
The statistical errors ($\chi^2$, yellow contour lines) yield 
uncertainties of $\pm 0.025$ in $Log(g)$ and $\pm 250$~K in $T_{\rm wd}$. 
The uncertainty in the Gaia distance (triple dashed blue line) introduces uncertainties of 
$\pm 0.02$ in $Log(g)$ and $\pm 100$~K in $T_{\rm wd}$.  
The systematic uncertainties in the STIS data/instrument are of only $\sim$3\%, and
introduce errors of the same order of magnitude as for the
distance: $\sim \pm 0.02$ in $Log(g)$ and $\sim \pm 100$~K in $T_{\rm wd}$.  

The uncertainties of the parameters due to uncertainty in the reddening/extinction law
are similar to what we obtain for TU Men: the reddening is the same and the distance
(dashed) line intersect the best-fit gray diagonal forming the same angle. 
For an uncertainty of 0.02 in the reddening $E(B-V)=0.08$, we have uncertainties
of $\pm 250$~K in $T_{\rm wd}$ and $\pm 0.15$ in $Log(g)$.  
All the values are listed in Table 4. 
 
Since the models are built in steps of 500~K in $T_{\rm wd}$ and 0.1 in $Log(g)$, 
the best-fit model we derive, even if all systematic and statistical errors 
were zero, can be off by {\it up to} 250~K and $0.05$ in $Log(g)$.  
Namely, there is an intrinsic uncertainty (systematic error) of maximum size 
$\pm 250$~K in $T_{\rm wd}$ and $\pm 0.05$ in $Log(g)$. 
Here too, if we linearly add all the uncertainties, 
overall we have  $Log(g)=8.275 \pm 0.25$ with $T_{\rm wd} = 30,000 \pm 1000$ K.

\clearpage

\subsubsection{{\bf The FUSE Spectral Analysis.}} 

For the FUSE spectrum of SS Aur,
we mask all the emission lines from the source 
(C\,{\sc iii} (977), O\,{\sc vi} ($\sim$1140)) as well as the sharp 
airglow/geocoronal emission lines. We mask all the region $\lambda < 950$ \AA , 
since the flux in that region also takes negative values (unreliable 
segments). We also mask all the sharp ISM molecular hydrogen lines. 
As the C\,{\sc iii} (1175) absorption line is absent, we also mask that
region. 

\begin{figure}[h!]
\vspace{-19.cm} 
\plotone{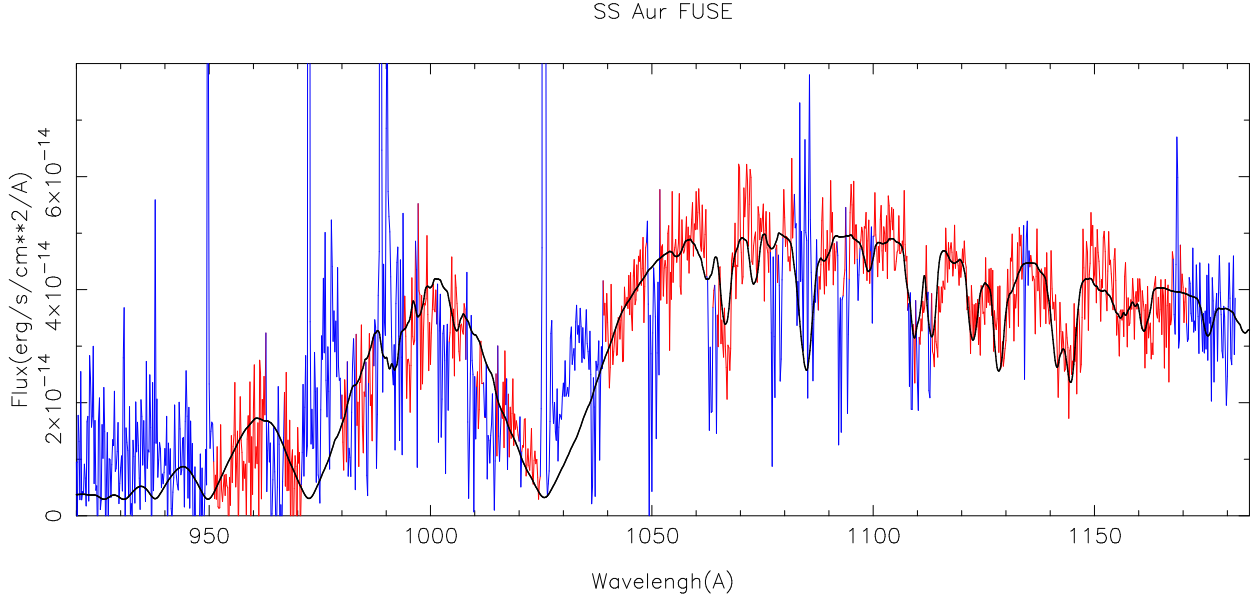} 
\vspace{-5.cm} 
\caption{
A WD model fit (black line) 
to the FUSE spectrum of SS Aur (red line) obtained
at quiescence on 2002 Feb 13, 28 days after outburst. 
The spectrum has been dereddened assuming E(B-V)=0.08. 
Before the spectral fit the sharp ISM absorption lines,
the sharp geocoronal emission lines, as well as 
the broad emission lines from the O\,{\sc vi} doublet ($\sim 1032, 1038$ \AA ),
and C\,{\sc iii} (977 \AA ), not forming in the photosphere,
are all masked and have been colored in blue.
Shortward of 950~\AA\  the flux is unreliable (it even takes some 
negative values) and has also been masked.  
The WD fit (Fig.\ref{ssaur_chi}) yields  to 
$T_{\rm wd}=33,375 $ K, 
for $Log(g)=8.66$ ($M_{\rm wd} = 1.045 M_{\odot}$). 
The fitting of the lines is given in section 6, but this model 
has a projected stellar rotational velocity $V_{\rm rot} sin(i) = 400 $ km/s   
with a carbon abundance [C]=0.001$\pm$0.001 solar. 
All the other species were set to solar abundance.    
\label{ssaurfuse} 
}
\end{figure}

The results for the $\chi^2$ values 
are presented in Fig.\ref{ssaur_chi}, also for $E(B-V)=0.08$, right panel. 
The best-fit model has 
$Log(g)=8.66$  with  a WD temperature $T_{\rm wd}=33,375$~K,
it is presented in Fig.\ref{ssaurfuse} displaying the
masked regions of the spectrum.

\clearpage

The uncertainties in the
FUSE data can reach up to 15\% in the continuum flux level,  
suffering from a measured flux that is to high 
by 5-10\% in one channel and off by up to 10\% in another channel. 
Whereas one does not expect more than a 5\% flux variation between  
FUSE and STIS spectra (see Sec.3), we assume now that the 
FUSE flux level is too high by 15\% (maximum error) and multiplied the 
FUSE spectrum
by a factor 0.85 before the spectral fit. The result is identical 
to the one presented in Fig.\ref{ssaur_chi}  but with the  
distance increased from 260~pc to 282~pc. Therefore, instead
of redrawing a new figure, we draw the $d=282$~pc distance line
in green in Fig.\ref{ssaur_chi}.   
This lowers the gravity by $\sim$0.15 in $Log(g)$ (from 8.66 to 8.50), 
and the temperature by $\sim$500~K (from 33,375~K to 32,875~K). 
In other terms, systematic errors in the FUSE instrument 
of $\pm 15$\% adds an error of $\pm 0.15$ in $Log(g)$ and 
$\pm 500$~K in $T_{\rm wd}$.    

Here too, an uncertainty of 0.02 in the reddening $E(B-V)=0.08$, 
gives  uncertainties of $\pm 250$~K in $T_{\rm wd}$ and $\pm 0.15$ in $Log(g)$.  
We also assume an intrinsic uncertainty of maximum size 
$\pm 250$~K in $T_{\rm wd}$ and $\pm 0.05$ in $Log(g)$
due to our {\it models finite size}. The statistical errors
gives the largest error in $T_{\rm wd}$ of $\pm 775$~K,
with $\pm 0.06$ in $Log(g)$. 
All are listed in Table \ref{errors}. 
The sum of the errors is pretty large and yields
$Log(g)=8.66 \pm 0.43$  with  $T_{\rm wd}=33,375 \pm 1875$~K. 
The large error on the temperature is due to the statistical
error, which we adopted for a 99\% confidence level, and the
instrument/calibration errors. We could have adopted a 90\%
confidence level together with a 5\% only of FUSE instrumental
errors, which would reduce the total error to $\Delta T \sim 1300$~K,
more in line with the STIS spectral analysis. 
The same is true for the error on $Log(g)$, which would be
reduced to $\sim 0.31$.    However, we chose the larger errors  
as we wish to compare the results from the FUSE analysis to that
of the STIS analysis (see next subsection).

\subsubsection{{\bf Comparison.}}  

Since the two (FUSE and STIS) spectra of SS Aur were not obtained
at the same epoch after the same outburst, we do not expect the temperature
to be the same, but we do expect the gravity to be the same. 
In order to compare the gravity obtained from the STIS analysis with 
that obtained from the FUSE
analysis, we do not include the errors propagating from the distance
and reddening errors, since they must be   
the same for the STIS and FUSE spectra. Taking into account only the statistical
errors, the instrument/calibration errors and the uncertainties 
from our own modeling (finite steps in $T_{\rm wd}$ and $Log(g)$), 
one has 
$T_{\rm wd}=30,000 \pm 600$~K with $Log(g)=8.28 \pm 0.10$ for STIS, and   
$T_{\rm wd}=33,375 \pm 1525$~K with $Log(g)=8.66 \pm 0.26$ for FUSE.   
The two gravity values {\it{ do not agree}}  within the error bars.
Though the overlap is only missed by a pretty small value, 0.02, 
it cannot be ignored. This is despite the fact that we chose relatively
large values for the FUSE instrument errors and for the statistical error. 

A possible explanation for the FUSE higher gravity is the higher temperature
itself: if the overall temperature was lower, it would move the gray diagonal
in Fig.\ref{tumenchi} higher up, which would then intersect the distance line at a lower
gravity value. We suspect that the  
possible presence of a hot (second) component contributing to the short
wavelengths of FUSE might be responsible for this discrepancy. 
We do not model such a second component as it 
would make the FUSE analysis diverge, since
the nature and spectral characteristics
of this component are completely unknown. In addition, there is no indication of such
a component in the STIS spectrum, which may indicate a transient
phenomenon. This unknown second emitting component introduces an
additional unknown systematic uncertainty which we do not take into account.  


\clearpage

\section{{\bf Abundances and Stellar Rotational Velocity.} } 

\subsection{{\bf TU Men}} 

We now model the abundance of the elements, starting 
with TU Men. 
We first generate a WD
model with $Log(g)=8.25$, and $T_{\rm wd}=27,750$~K, and 
vary the elemental abundances one by one. For each value of
an elemental abundance we vary the projected rotational stellar
velocity as explained in Sec.4.3.  
For the STIS spectrum of TU Men, we model the abundances
of carbon, nitrogen and silicon.  In order to fit the wavelength 
of (most of) the carbon and silicon lines,
the model spectra had to be (red) shifted by +0.6~\AA . 

As an example of how we derive abundances and broadening velocity 
by visual examination, we explicitly show the fitting of the 
Si\,{\sc ii}/Si\,{\sc iii} ($\sim 1300$~\AA ) absorption lines 
in Fig.\ref{tumen_silicon}.  
From this figure we obtain [Si]=$0.2 \pm 0.1 \times $ solar with a 
velocity broadening of $250 \pm 50$~km/s. 
For this abundance 
(second row from the top) the depth
of the complex silicon spectral feature in the model (red line) 
is the same as in the observed spectrum (black line). If we 
consider higher abundances (lower rows) we clearly see how the spectral
feature in the model becomes deeper  than in the observed spectrum. 
In that regard, the $\chi^2_{\nu}$ technique agrees with the
visual examination as shown in the left panel of Fig.\ref{tumensierrors}: 
the least $\chi^2_{\nu}$ fit to the silicon $\sim$1300~\AA\ region 
is minimum for [Si]=0.2$\times$solar.  
The small flux fluctuations as well as possibly some shallow and sharp
absorption lines in the spectrum are of the same amplitude 
as the flux errors.  As a consequence, 
the best-fit broadening velocity of 250~km/s 
is found by looking at the 3 individual lines in the silicon feature, 
as shown with arrows in the right panel of Fig.\ref{tumensierrors}.   
These 3 lines have a depth that is larger than the amplitude of
the errors and are braoder than the spectral binning of $0.58$\AA .  
While the visual examination clearly shows that the best-fit is 
for [Si]$\approx 0.1-0.2$ for $V=250$~km/s, the $\chi^2$ becomes smaller
for the higher velocity models,  $V\sim 350$~km/s and higher, 
see left panel of Fig.\ref{tumensierrors}.  
The reason the $\chi^2_{\nu}$ minimization technique gives a much higher
velocity is that it cannot identify actual lines and 
differentiate them from the small fluctuations. 
As the velocity broadening increases, the $\chi^2_{\nu}$ decreases 
sharply until it reaches about 300~km/s, at higher velocities
the $\chi^2_{\nu}$ decreases only by a few percent at most, which
is not significant enough to assess which of these models, 
say e.g. V=300~km/s vs. V=600~km/s, is the best.      
The important point, however, is that, irrelevant to the broadening
velocity (250~km/s vs. 450~km/s), a best fit of [Si]=0.2 is obtained
for both the $\chi^2_{\nu}$ technique and the visual examination.   
It becomes clear that for a higher abundance and higher velocity, e.g. 
[Si]=0.5 with V=350~km/s - the lower right panel in Fig.\ref{tumen_silicon}, 
there is a fairly gross discrepancy between the model (red line) 
and data (black line): a higher velocity with a higher abundance
degrades the fit.

\clearpage

\begin{figure}[h!]
\gridline{
          \fig{tuSi0.1v150.eps}{0.115\textwidth}{[Si]=0.1,~V=150~km/s}
          \fig{tuSi0.1v200.eps}{0.115\textwidth}{[Si]=0.1,~V=200~km/s}
          \fig{tuSi0.1v250.eps}{0.115\textwidth}{[Si]=0.1,~V=250~km/s}
          \fig{tuSi0.1v300.eps}{0.115\textwidth}{[Si]=0.1,~V=300~km/s}
          \fig{tuSi0.1v350.eps}{0.115\textwidth}{[Si]=0.1,~V=350~km/s}
}
\vspace{-1.5cm} 
\gridline{
          \fig{tuSi0.2v150.eps}{0.115\textwidth}{[Si]=0.2}
          \fig{tuSi0.2v200.eps}{0.115\textwidth}{}
          \fig{tuSi0.2v250.eps}{0.115\textwidth}{}
          \fig{tuSi0.2v300.eps}{0.115\textwidth}{}
          \fig{tuSi0.2v350.eps}{0.115\textwidth}{}
}
\vspace{-1.5cm} 
\gridline{
          \fig{tuSi0.3v150.eps}{0.115\textwidth}{[Si]=0.3}
          \fig{tuSi0.3v200.eps}{0.115\textwidth}{}
          \fig{tuSi0.3v250.eps}{0.115\textwidth}{}
          \fig{tuSi0.3v300.eps}{0.115\textwidth}{}
          \fig{tuSi0.3v350.eps}{0.115\textwidth}{}
}
\vspace{-1.5cm} 
\gridline{
          \fig{tuSi0.4v150.eps}{0.115\textwidth}{[Si]=0.4}
          \fig{tuSi0.4v200.eps}{0.115\textwidth}{}
          \fig{tuSi0.4v250.eps}{0.115\textwidth}{}
          \fig{tuSi0.4v300.eps}{0.115\textwidth}{}
          \fig{tuSi0.4v350.eps}{0.115\textwidth}{}
}
\vspace{-1.5cm} 
\gridline{
          \fig{tuSi0.5v150.eps}{0.115\textwidth}{[Si]=0.5}
          \fig{tuSi0.5v200.eps}{0.115\textwidth}{}
          \fig{tuSi0.5v250.eps}{0.115\textwidth}{}
          \fig{tuSi0.5v300.eps}{0.115\textwidth}{}
          \fig{tuSi0.5v350.eps}{0.115\textwidth}{}
}
\caption{
Example of fitting absorption lines by visual examination: 
the fitting of the Si\,{\sc ii} \& Si\,{\sc iii} ($\sim$ 1300~\AA ) 
absorption feature in the STIS spectrum of TU Men is shown explicitly.  
The STIS spectrum is in black, the WD model is in red with a 
temperature of 27,750~K and an effective surface gravity of
$Log(g)=8.25$. Each of the 25 panels displays the same  
wavelength coverage: from 1280~\AA\ to 1320~\AA . 
The silicon abundance increases from 0.1 solar (top row) to 0.5 solar
(bottom row) in steps of 0.1; the velocity broadening of the lines is 
increased from 150~km/s (far left) to 350~km/s (far right) in steps 
of 50~km/s. 
A visual inspection of the panels immediately shows that the 
150~km/s and 350~km/s broadening velocities can be excluded.
Similarly it is clear that the abundances [Si]=0.4 and 0.5 solar 
yield to deeper absorption features (which is more apparent when
looking at the 350~km/s broadening). The best-fit is 
for [Si]$\approx 0.1-0.2$ solar with $V=250$~km/s. Due to the
finite size of our grid of models we adopt the solution
[Si]= $0.2 \pm 0.1 $ solar with $V=250 \pm 50$~km/s. 
Note that the $V=50$, 100, 400, 450, and 500~km/s, as well as the
abundances [Si]=0.6, 0.7, 0.8, 0.9, and 1.0 are not shown for clarity.    
\label{tumen_silicon} 
}
\end{figure} 

\clearpage 

\begin{figure} 
\plottwo{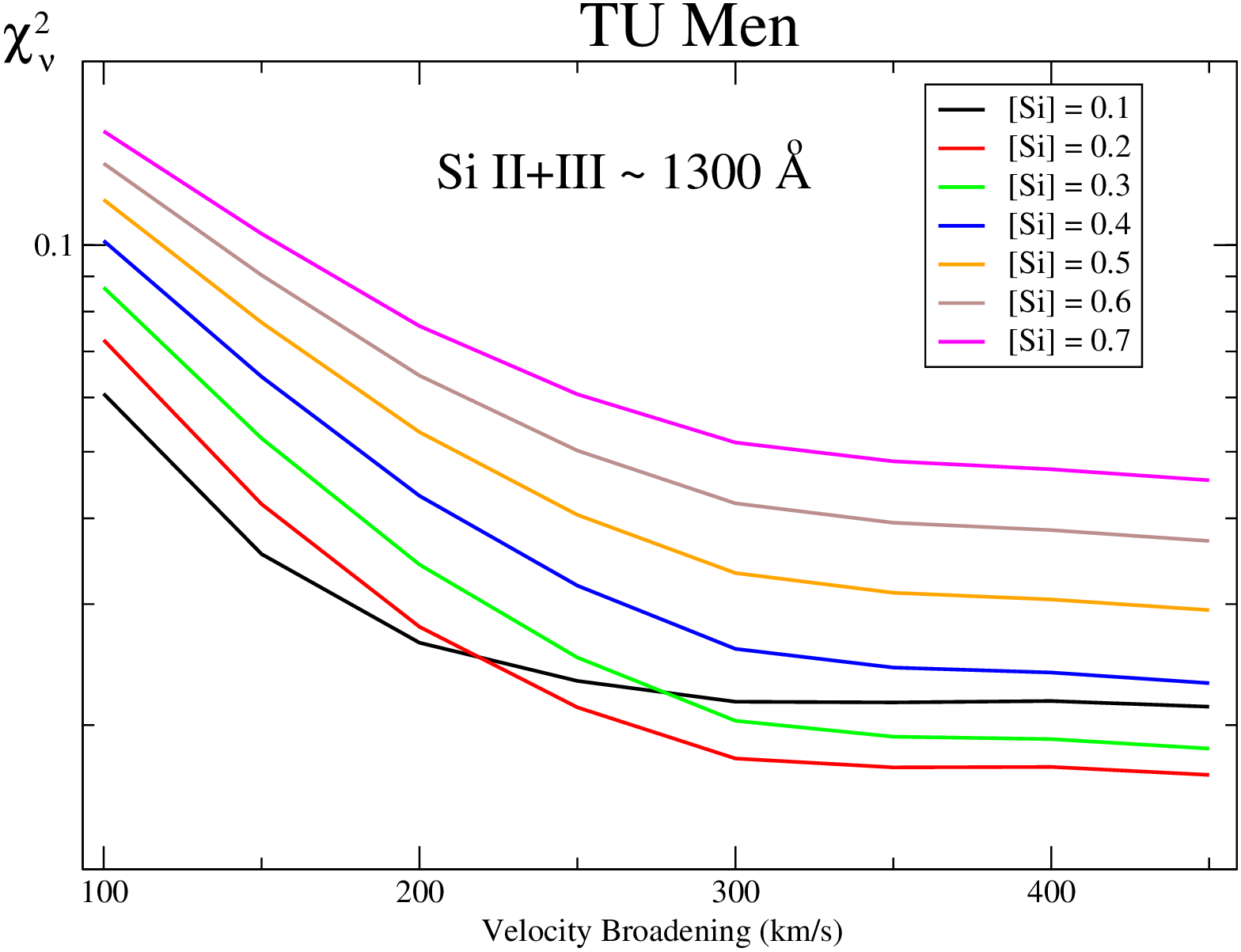}{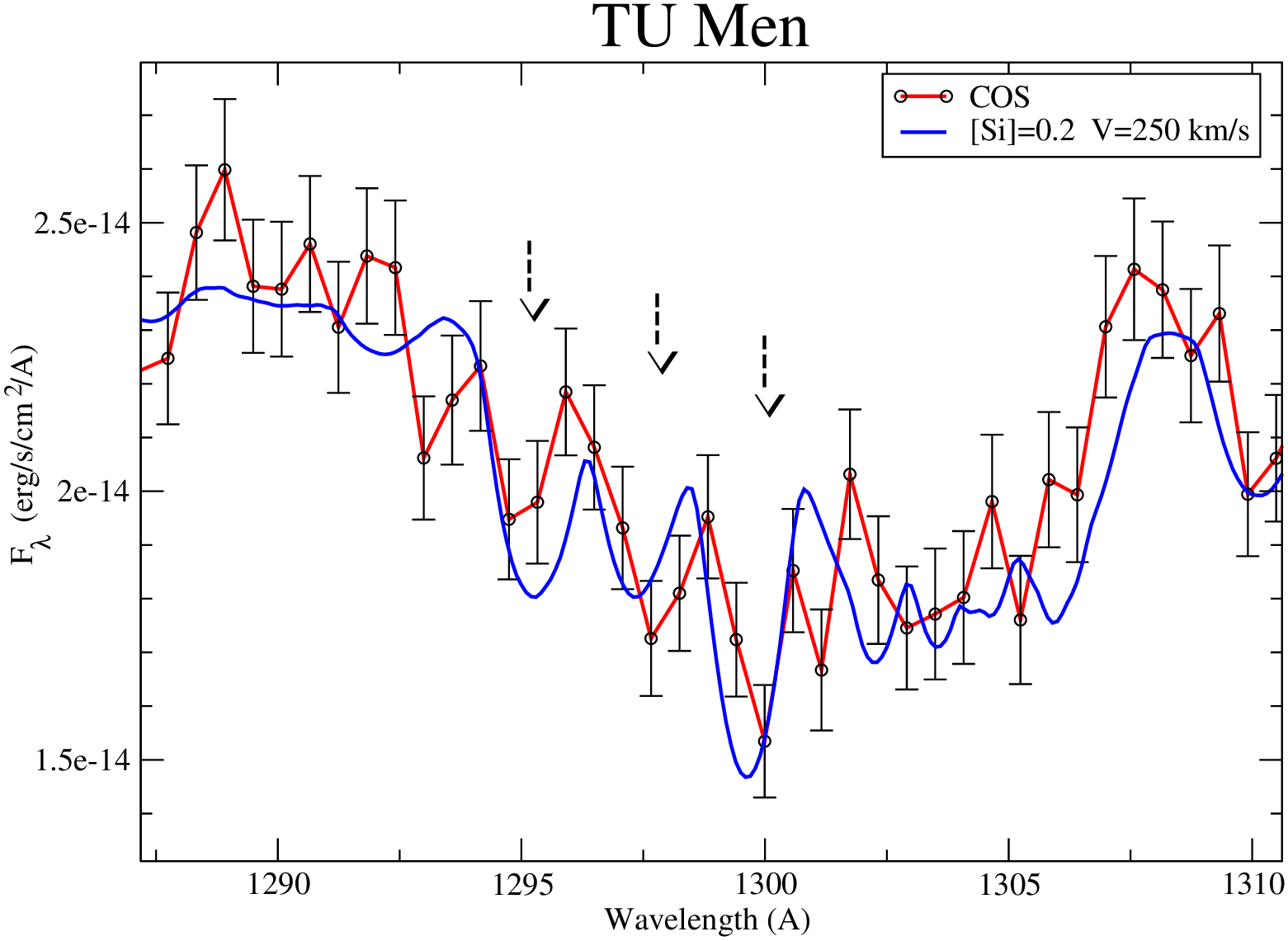}
\caption{
Fitting the Si\,{\sc ii} + Si\,{\sc iii} absorption lines
at $\sim 1300$~\AA\ (as shown in Fig.\ref{tumen_silicon}) yields
to a low $\chi^2_{\nu}$ best-fit for the largest values of the broadening
velocity, as shown on the  {\bf left.} This is due to the spectral 
binning (0.58~\AA ) and errors having the same width and amplitude
as the small flux fluctuation of the spectrum, as shown on the {\bf right},
without being able to differentiate between fluctuations and actual 
absorption lines. 
The visual examination (done in Fig.\ref{tumen_silicon}) is carried 
out by ensuring that actual absorption lines, as the ones indicated
by the three arrows in the right panel, are matched in the fitting
(within the given accuracy of the flux errors and spectral binning
size).  
In spite of that the $\chi^2_{\nu}$ minimization technique does 
agree with the visual examination that the silicon abundance 
is small: [Si]=0.2.  At higher broadening
velocity the [Si]=0.2 model does not reproduce these features in 
Fig.\ref{tumen_silicon}. A higher velocity with a higher abundance
also does not provide a fit visually good enough (as can also 
be seen in the previous figure).  
\label{tumensierrors} 
}
\end{figure} 

\clearpage 

\begin{figure}[h!]
\vspace{-2.cm} 
\gridline{
          \fig{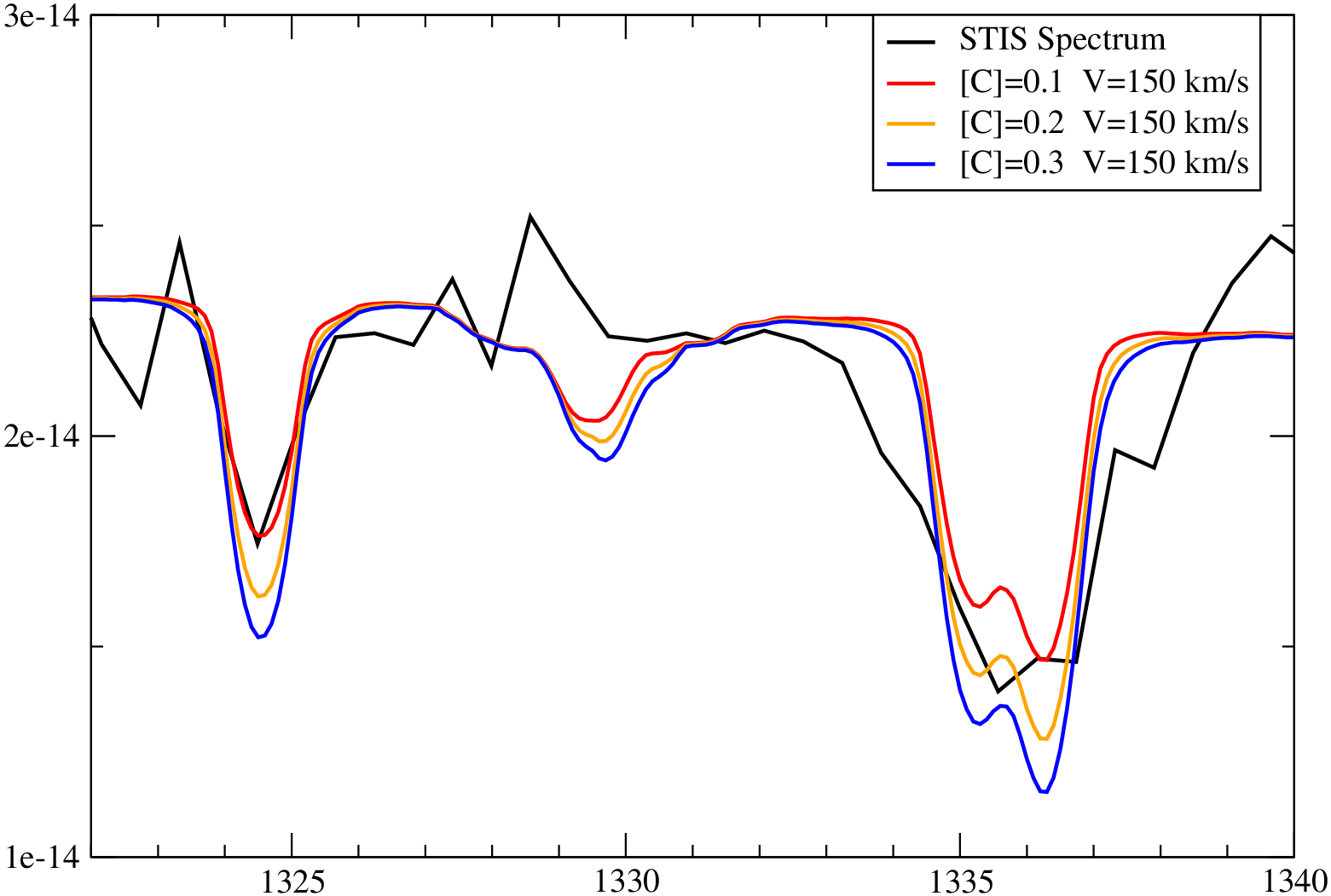}{0.30\textwidth}{}
          \fig{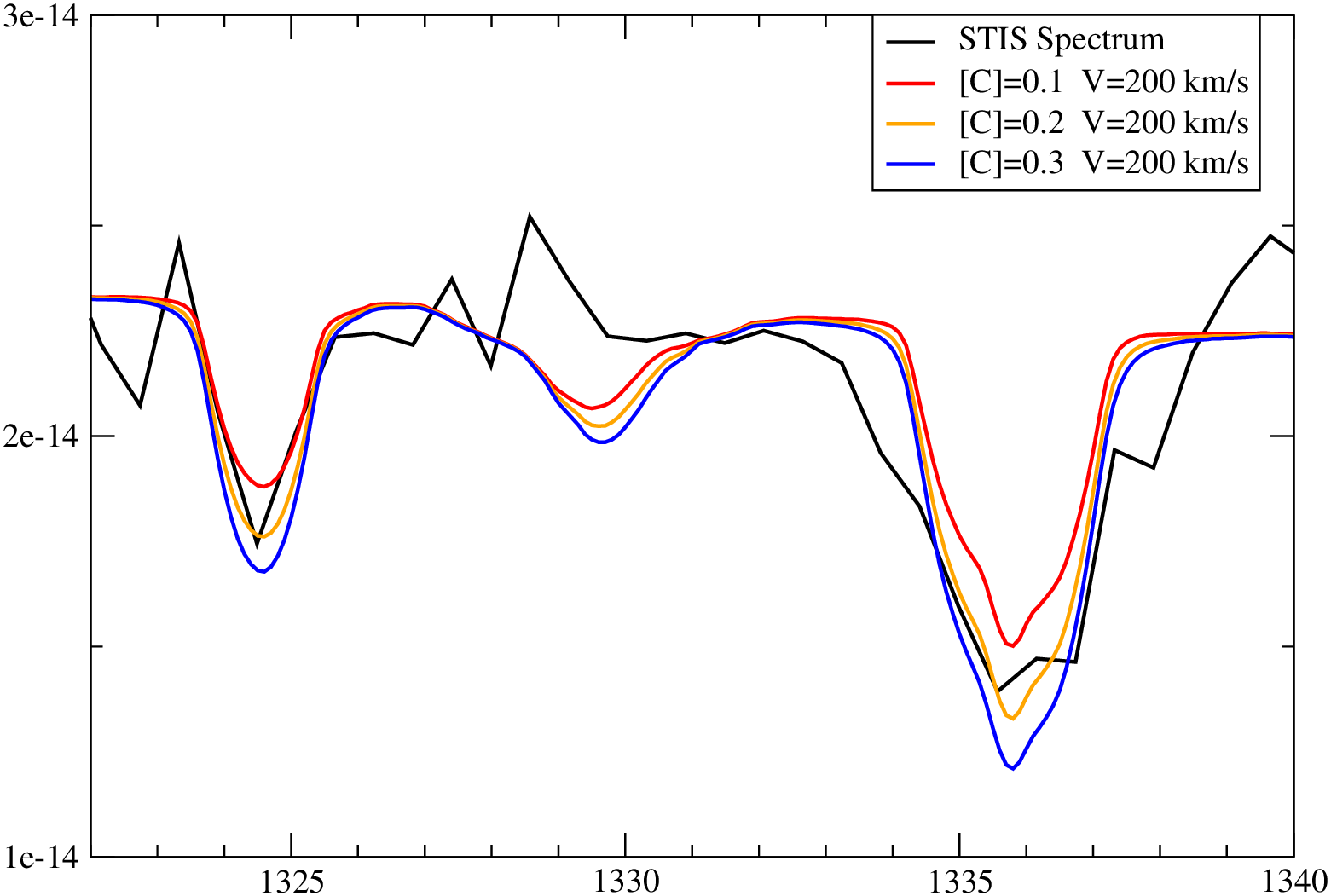}{0.30\textwidth}{}
          \fig{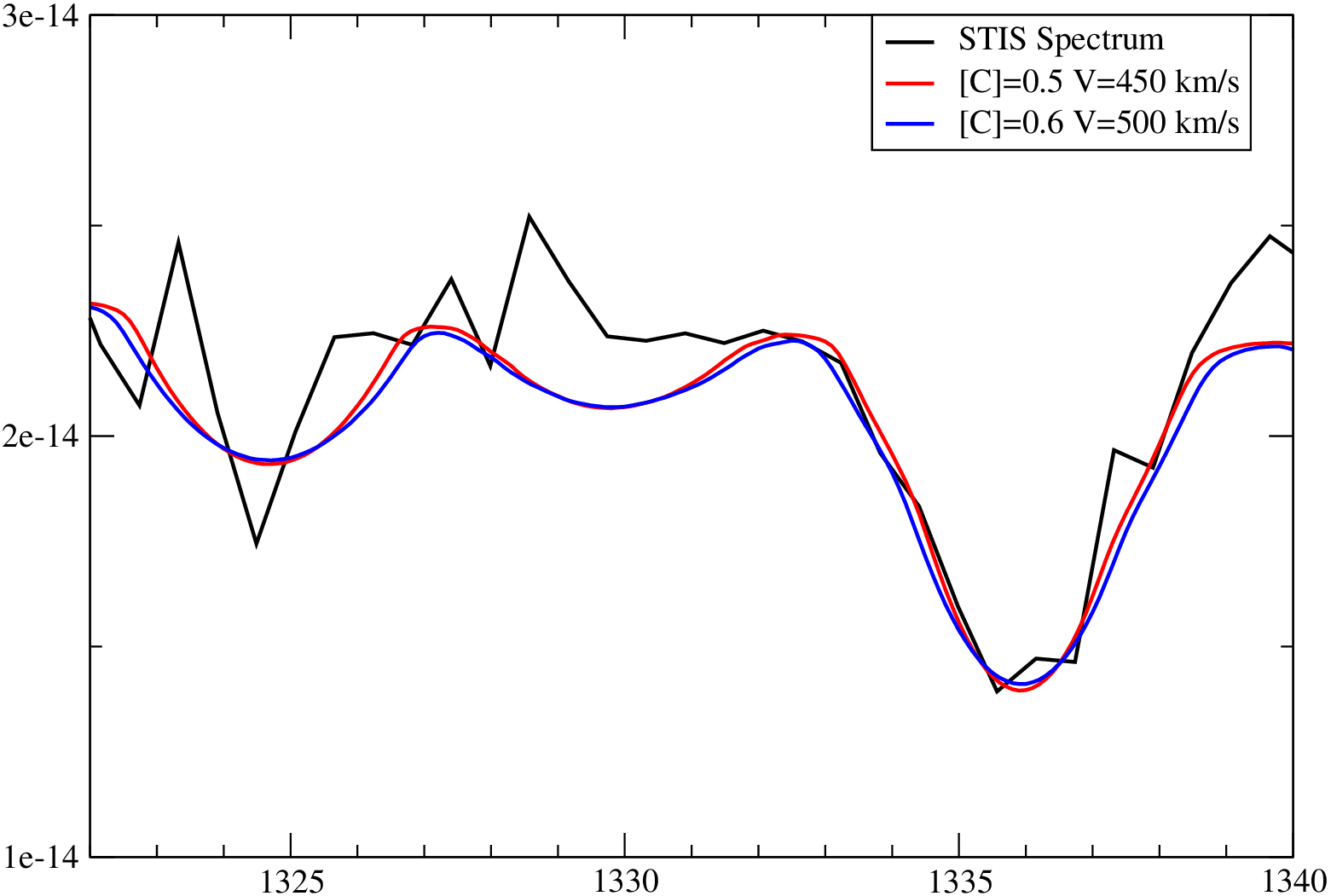}{0.30\textwidth}{}
}
\vspace{-1.cm} 
\caption{
The fitting of the C\,{\sc ii} ($\sim$1325 \& $\sim$1336) 
absorption lines in the STIS spectrum of TU Men is shown as a function
of the carbon abundance [C] and velocity broadening $V$. 
The C\,{\sc ii} (1325) line is best fitted with [C]=0.1 and  
$V=150$~km/s (left panel). The C\,{\sc ii} (1336) doublet agrees better
with [C]=0.2 and $V=200$~km/s (middle panel), when one fits the lower 
narrower part of the absorption feature. The broader (and shallower) part of  
the feature agrees with [C]=0.5-0.6 and $V=450-500$~km/s (right panel). 
Such a large velocity broadening is in disagreement with all the other 
absorption lines,  raising the possibility that the broadening is due 
to a second component.  
\label{tumen_carbon} 
}
\end{figure}

We also illustrate, in Fig.\ref{tumen_carbon}, 
how we fit the carbon abundance based mainly
on the C\,{\sc ii} ($\sim 1325$ \& $\sim 1336$) absorption lines. 
The C\,{\sc ii} ($1325$) absorption line agrees best with 
[C]=$0.1 $ (solar) and $V=150$~km/s,
and somewhat also with [C]=$0.2 $ and $V=200$~km/s. 
As to the C\,{\sc ii} ($1336$) line, its bottom agrees well  
with [C]=$0.2 $  and $V=200$~km/s, 
and somewhat also with [C]=$0.3 $ and $V=250$~km/s. 
The shallower part, however, can be fitted with [C]=$0.5-0.6 $ with 
$V=450-500$~km/s. The two different widths may indicate a second
component polluting the line, such as e.g. ISM absorption. 
Indeed, the local ISM (within 100 pc) could contribute to the 
C\,{\sc ii} (1336) line (as well as to the Si\,{\sc ii} (1260) line)
\citep{red04}, and the shape of the C\,{\sc ii} (1336) line indicates 
a possible (less deep) second component.  
Therefore, for carbon, we have [C]=$0.2 \pm 0.1$ with 
$V=200 \pm 50$~km/s.  
In a similar manner we fit the nitrogen lines N\,{\sc i} 
($\sim 1165-1170$).

\clearpage

\begin{figure}[h!]
\vspace{-3.cm} 
\gridline{\fig{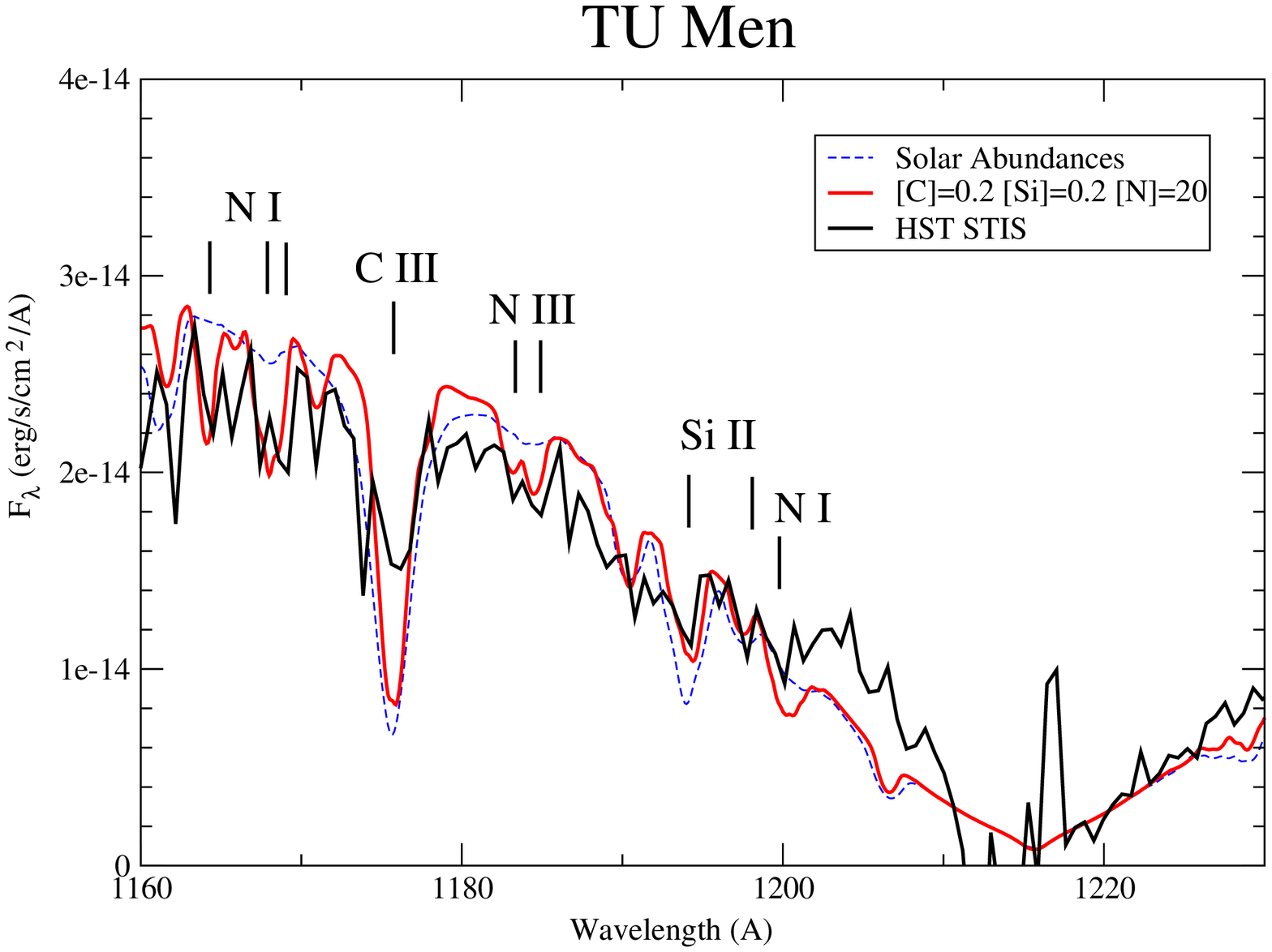}{0.44\textwidth}{}
          \fig{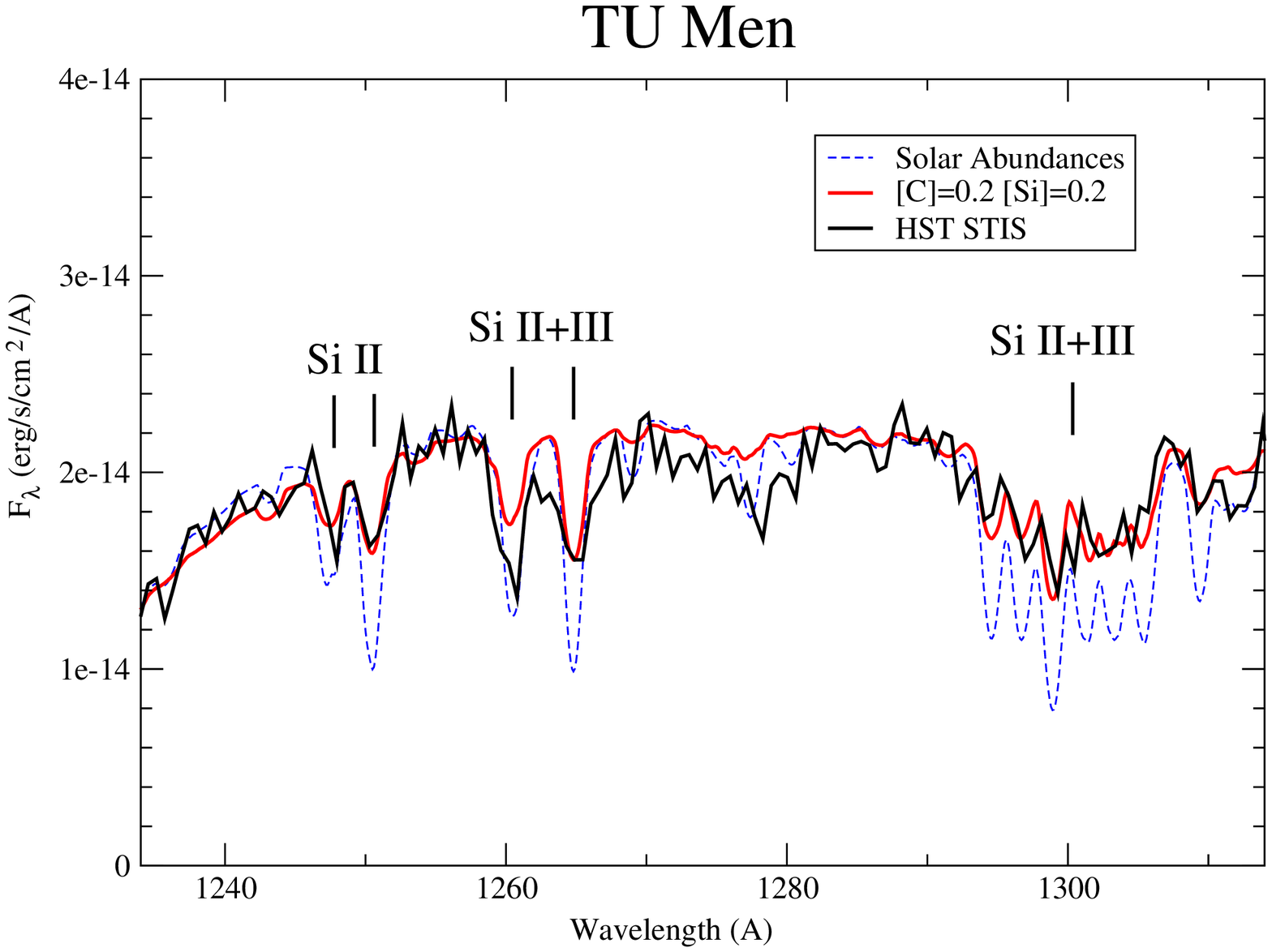}{0.44\textwidth}{}
}
\vspace{-1.cm} 
\caption{
The fits to carbon, silicon, and nitrogen absorption lines are shown
in details.  
The HST spectrum is in black, the model is shown in red, and for 
comparison a solar abundance model is shown with the blue dashed line. 
The left panel shows the blue wing of the Lyman $\alpha$ region 
and the fitting of (mainly) the nitrogen lines; 
the right panel shows the fitting of silicon lines.  
The abundances are as indicated in the upper right in each
panel. 
The C\,{\sc iii} (1175) may have emission at the bottom of its
absorption line. 
\label{tumenab} 
}
\end{figure}

The best fit to the absorption lines gives: 
[C]=$0.2 \pm 0.1$ solar,  
[Si]=$0.2 \pm 0.1$ solar, 
and [N]=$20 \pm 10$ solar, 
with an overall matching broadening velocity of $\sim 225 \pm 75$~km/s.  
The final result of this model (temperature, gravity, abundances
and velocity broadening) was presented in Fig.\ref{tumenfit}. 
While fitting of the absorption lines is further presented in 
the 2 panels of Fig.\ref{tumenab}, for wavelengths below 1280~\AA\   
(since the remaining lines are already presented in Figs.\ref{tumen_silicon}
\& \ref{tumen_carbon}).
For comparison, in Fig.\ref{tumenab}, we also show a solar abundance model.
The C\,{\sc iii} $\lambda \sim$1175  line seems
to possibly have a blue shifted emission peak
in its center (first panel). 
While we fit most of the silicon lines,
the Si\,{\sc ii} (1260) line is deeper than in our model. 
From the shape of the line (showing a sharp bottom) and from the
above mention ISM absorption, it is likely that this line is also
affected by ISM absorption.

We also check how the error on the temperature and gravity affect
the error on the abundances (propagation of errors).  
We compare the absorption lines of carbon [C]=0.2, 
silicon [Si]=0.2, and nitrogen [N]=20 for the two limiting
cases (within the error bars, see Table 5): 
$T_{\rm wd}=26,750$~K with $Log(g)=8.00$, against   
$T_{\rm wd}=28,750$~K with $Log(g)=8.50$. We find that the
difference in the absorption lines profiles between these  
two models is significantly smaller than the 
size of the abundance steps in the models. The propagation of the errors in $T_{\rm wd}$
and $Log(g)$  is negligible compared to the errors we adopted for
the abundances.

\clearpage

\subsection{{\bf SS Aur}} 

To fit the abundances in  
the STIS spectrum of SS Aur, we set $T_{\rm wd} = 31,000$~K with  
$Log(g)=8.50$, this corresponds to the upper values of the STIS results
within the errors (see Table 5). 
For SS Aur, we fit the abundances 
of silicon and carbon, following the same procedure described
for TU Men, namely, we vary the abundance of Si and C one by one, and for each
abundance value, we vary broadening velocity (as already explained and 
illustrated in Fig.\ref{tumen_silicon}).  

\begin{figure}[b!] 
\vspace{-9.cm} 
\plottwo{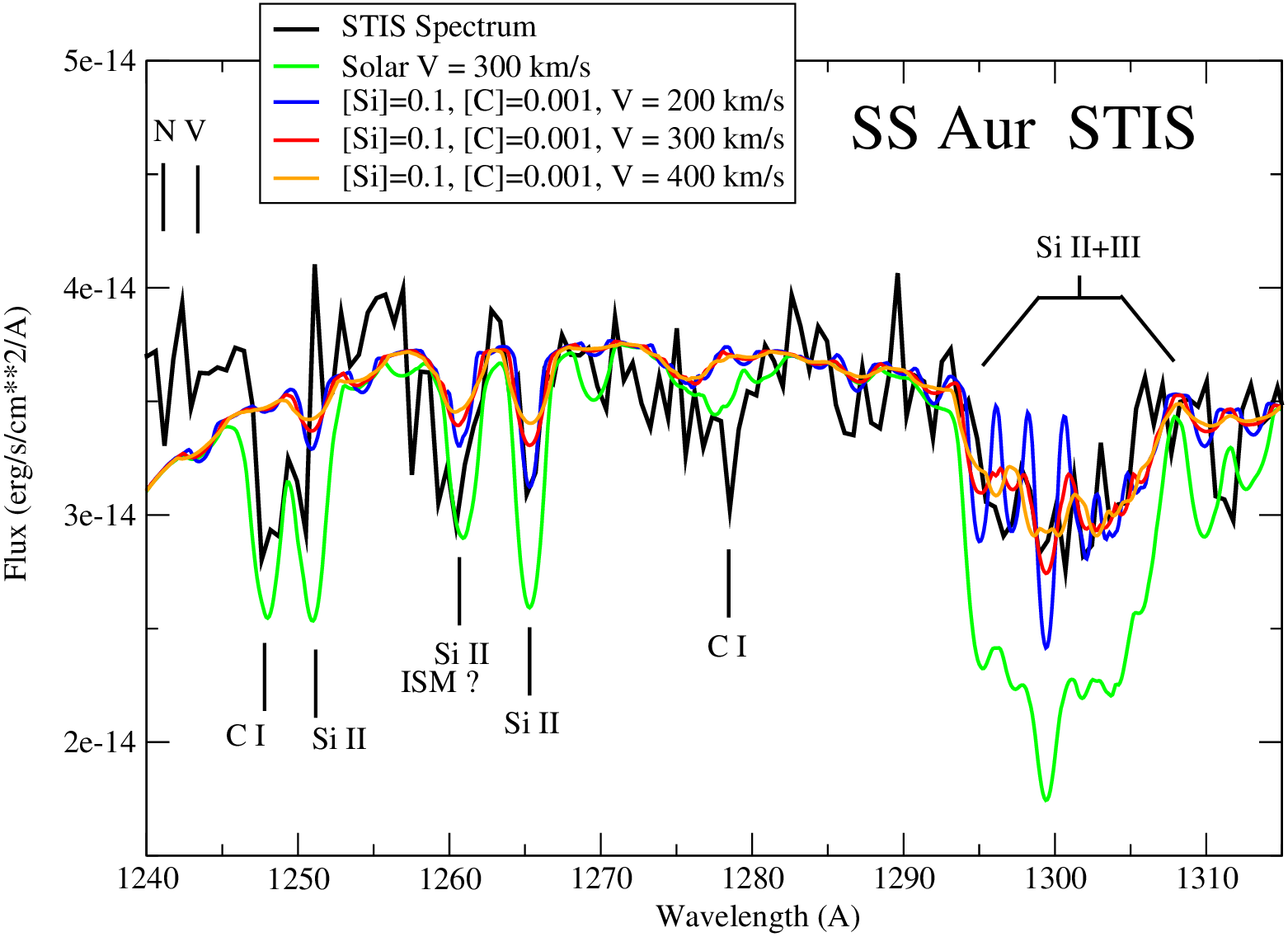}{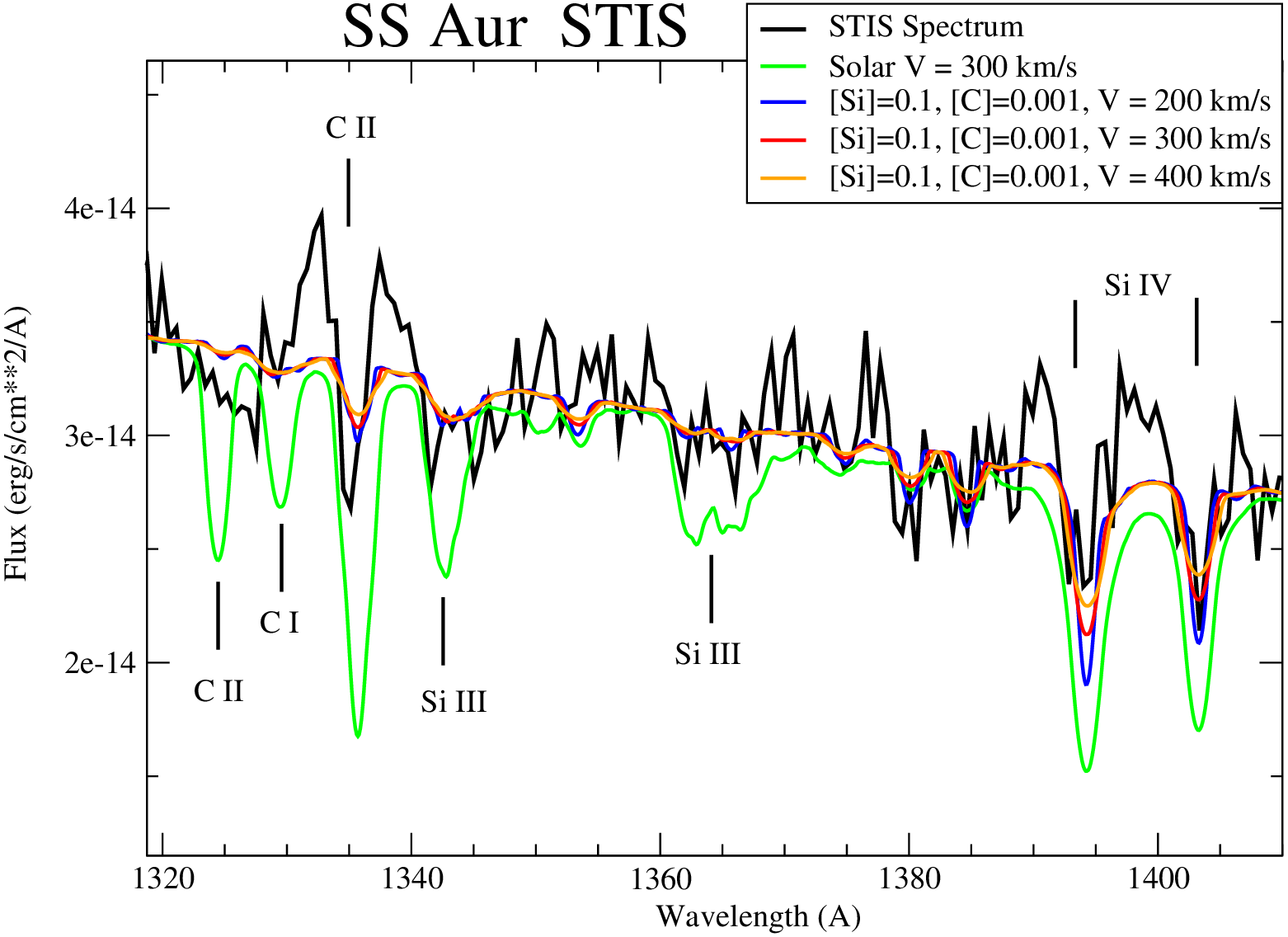} 
\caption{ 
Detailed regions of the spectral fit are displayed, showing
different model fits to the STIS spectrum (in black).
The model has $T_{\rm wd}=31,000$~K with $Log(g)=8.5$, the abundances
and broadening velocities are as indicated (color-coded).  
The fit to the absorption lines is discussed in the text.  
\label{ssaurstisab} 
 }
\end{figure}

The fitting of the lines in the STIS spectrum of SS Aur 
is illustrated in Fig.\ref{ssaurstisab}. 
We display a solar abundance model (green), and a low silicon
abundance model with different broadening velocity
$V=200$~km/s (blue), $V=300$~km/s (red), and $V=400$~km/s (orange).   
We note (left panel) that the Si\,{\sc i} ($\sim$1265) line agrees with 
$V=200$~km/s, while the Si\,{\sc ii}+Si\,{\sc iii} complex absorption
feature ($\sim 1300$) agrees with $V=300$ \& 400~km/s. 
As for TU Men, the Si\,{\sc i} (1260) is possibly from the ISM. 
The C\,{\sc i} \& Si\,{\sc ii} lines ($\sim 1250$) are possibly
affected by some emission (N\,{\sc v} $\sim$1240), as the continuum
there doesn't match, but while the carbon could be solar, the silicon
line is definitely much sharper (ISM?). Both the C\,{\sc ii} ($\sim$1335)   
and Si\,{\sc iv} ($\sim$1400) lines also reveal broad emission
(right panel), but in both cases the absorption feature seems less
deep than solar. In addition the C\,{\sc ii} ($\sim$1325),   
C\,{\sc i} ($\sim$1330),  and Si\,{\sc iii} ($\sim$1343 \& $\sim$1365) 
all agree well with and a low silicon and carbon abundances. 
Overall, the fitting of the absorption lines is consistent with a 
very low carbon abundances($\sim 0.001$ solar), 
and agrees with a silicon abundances [Si]=$0.1^{+0.1}_{-0.05}$.
We assess a broadening velocity $V \sim 300 \pm 100$~km/s.  
 
We compare this abundance model with a similar one in which we 
set  $T_{\rm wd} = 29,000$ K with $Log(g)=8.00$.
These values of $T_{\rm wd}$ and $Log(g)$ correspond to the lower end   
of the STIS results for SS Aur within the errors. As for the TU Men, we find that 
the difference between the lines profiles of the two models (31,000~K and 29,000~K) 
is well within the error bars of the abundances (i.e. the size of the 
step by which we  increase the abundances). 
Namely, we obtain that the errors on the abundances propagating from the errors on 
$T_{\rm wd}$ and $Log(g)$ are negligible as they are smaller than the size of
the abundances step.

\clearpage

\begin{figure}[h!] 
\vspace{-9.0cm} 
\plotone{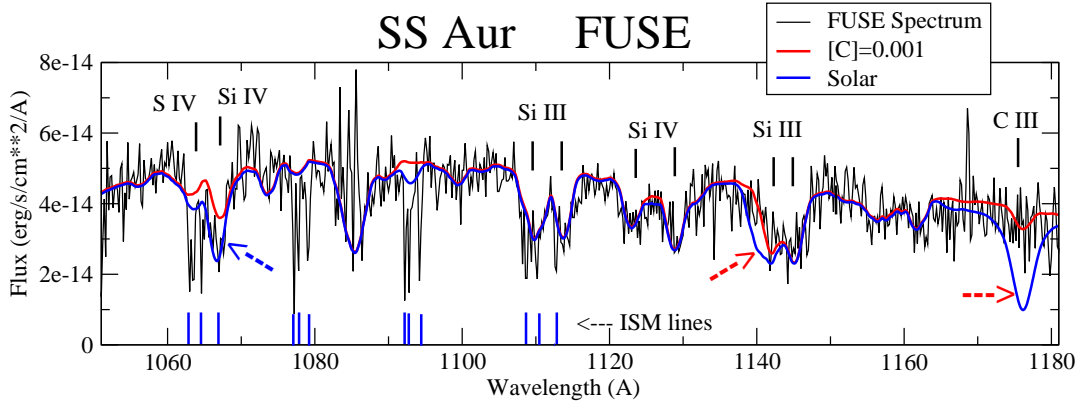} 
\vspace{-5.cm} 
\caption{
A detailed portion of the FUSE (in black) spectral fit is displayed
showing a solar abundance model (in blue) compared to a low carbon
abundane model (in red). In order to fit the 
C\,{\sc iii} (1175 \AA ) line (red arrows on the right),  the carbon abundance was lowered
to 0.001 (in solar units), while the remaining species were left
at their solar value. The low carbon also improves the fit 
to the C\,{\sc i} absorption feature near 1139 \AA\ (red arrow), but degrades
the fit to the C\,{\sc ii} feature near 1163-66 \AA\ (blue arrow on the left). 
That C\,{\sc ii} feature is also affected by ISM absorption.  
The model has $T_{\rm wd}=32,500$~K with $Log(g)=8.5$ with a rotational
velocity of 400~km/s.  
\label{ssaurfusedetail}  
}
\end{figure}

\begin{figure}
\vspace{-12.0cm} 
\plotone{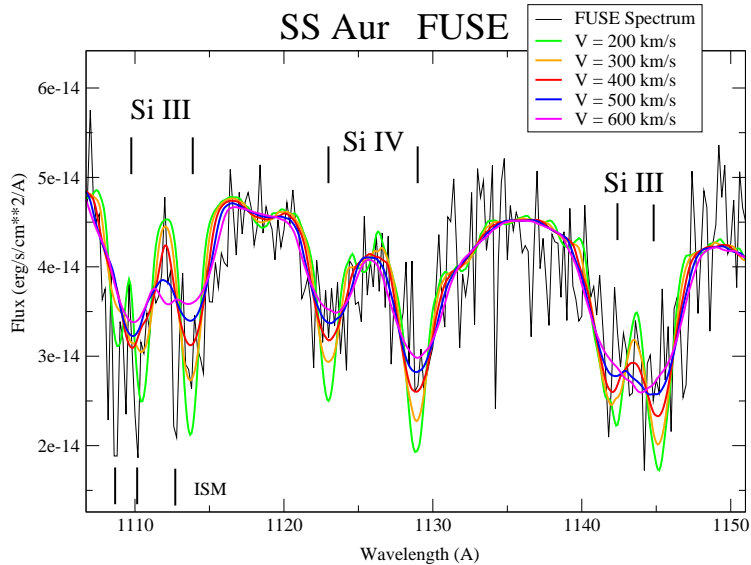}
\vspace{-0.5cm} 
\caption{
The region of the silicon absorption lines unaffected by ISM absorption 
(1120~\AA\ - 1150~\AA ) 
is shown for different broadening velocities, solar silicon abundance and low
(0.001) carbon abundance. The fitting gives a velocity of $500 \pm 100$~km/s. 
The ISM lines are marked at the bottom left of the panel. 
The model has $T_{\rm wd}=32,500$~K with $Log(g)=8.5$.  
\label{ssaurfusevel}
}
\end{figure}

For the abundance analysis of the FUSE spectrum of SS Aur, 
we set $T_{\rm wd} = 32,500$ K, with $Log(g)=8.5$, the same $Log(g)$
assumed for the STIS spectrum of SS Aur.  
The fitting of the
abundances yields to solar abundances, except for carbon, which
gives [C]=$0.001\pm0.001$. This is based mainly on the C\,{\sc iii} (1175) 
absorption line, but a detailed look at absorption lines 
(see Fig.\ref{ssaurfusedetail}) shows that another carbon absorption
line ($\sim$1140~\AA ) is also better fitted with a very low carbon
abundance. On the other hand, a carbon absorption feature near 1165~\AA\ 
seems to be better fitted with solar carbon abundance. That region, however,
is also affected by ISM absorption lines. 
For the projected stellar velocity we obtained $\sim 500 \pm 100$ km/s
based on the (unaffected by ISM) silicon absorption lines
(Fig.\ref{ssaurfusevel}).

\clearpage 

Next, we compute a model with $T_{\rm wd}=35,250$~K and $Log(g)=9.1$, i.e. the
upper limit of the FUSE solution (within its error bars) and a model 
with $T_{\rm wd}=31,500$~K and $Log(g)=8.2$, i.e. the
lower limit of the FUSE solution (also within its error bars, see Table 5). 
We set up solar abundances, except for [C]=0.001 solar. 
The hotter model exhibits slightly shallower Si\,{\sc iii} ($\sim$1145) lines 
and slightly deeper Si\,{\sc iii} ($\sim$1110) lines than the colder model, 
but here too the difference is negligible when compared to the step size of the 
the abundance. 
Namely, the propagation of the errors in temperature and gravity on the silicon 
and carbon abundance is negligible.

\section{{\bf Discussion and Conclusion}}

We summarize the results in Table \ref{results}, where we also include the
mass and radius of the WD as derived from the $Log(g)$ values using 
the mass-radius relation for non-zero WD \citep{woo95}. 
For SS Aur the FUSE analysis yielded a higher temperature and, therefore, 
a higher gravity than the STIS analysis. The difference in gravity was not within
the limits of our error analysis. A possible cause for the higher temperature, 
could be a second (hot) component contributing only to the shortest wavelength of
the FUSE spectrum. 

\begin{deluxetable}{lcccc}[h!]  
\tablewidth{0pt}
\tablecaption{FUV Spectral Analysis Results 
\label{results} 
} 
\tablehead{ 
Parameter    & Units         &        TU Men          &       SS Aur       &   SS Aur    \\      
             &               &        STIS            &       STIS         &   FUSE            
}
\startdata
$T_{\rm wd}$ & (K)           & $27,750 \pm 1000$     & $30,000\pm 1000$    & $33,375 \pm 1875$         \\[3pt]  
$Log(g)$     & (cgs)         & $8.25 \pm 0.25 $      & $8.275\pm0.25 $     & $8.66 \pm 0.43$           \\[3pt]  
{\bf $\chi^2_{\nu}$ } &             & 1.247                  & 1.324              & 0.959                  \\[3pt] 
$M_{\rm wd}$&$(M_{\odot})$   & $0.77^{+0.16}_{-0.13}$ & $0.785_{-0.13}^{+0.17}$ & $1.045_{-0.28}^{+0.21}$ \\[3pt]
$R_{\rm wd}$ & (km)          & $7550^{+1610}_{-1270}$&$7430^{+1620}_{-1290}$&$5490^{+2205}_{-1835}$   \\[3pt]   
$[$C$]$      & Solar         & $0.2 \pm 0.1$          & $<< 1$             & $<< 1$                   \\[3pt]  
$[$N$]$      & Solar         & $20 \pm 10$            & ---                &  ---                     \\[3pt]  
$[$Si$]$     & Solar         & $0.2 \pm 0.1$          & $0.1_{-0.05}^{+0.1}$ &    1                   \\[3pt] 
$V_{\rm rot} sin(i)$& (km/s) & $225\pm75$             & $300\pm 100$         & $500\pm100$            \\   
\enddata
\tablecomments{
The STIS and FUSE spectra of SS Aur were not obtained during the same epoch, as a consequence the abundances
and temperature of the WD in column 4 \& 5 are not expected to agree. The masses and radii were obtained 
from the values of $Log(g)$ using the mass-radius relation for non-zero temperature WD \citep{woo95}. 
Since for each spectrum the solution is a narrow diagonal band in the $T_{\rm wd}$ vs $Log(g)$ parameter space, 
the larger temperature (+) is associated with the larger gravity (+), larger WD mass (+), 
and smaller WD radius (-), and {\it vice versa}. 
} 
\end{deluxetable}

The dereddened STIS snapshots of SS Aur and TU Men were analyzed in \citet{sio08}, 
who considered WD models, disk models as well as WD+disk models, but the
disk only degraded the solution, confirming that the observed spectrum
is that of an exposed WD. Only solar composition WD models were fitted
to the STIS spectra assuming $Log(g)=8.0$ and $8.3$ for TU Men, 
and $Log(g)=8.7$ and $8.8$ for SS Aur. 
In \citet{god08} we analyzed the combined FUSE+STIS spectrum of SS Aur, 
considering values of $Log(g)=8.3$ to $8.9$ yielding a temperature 
$T_{\rm wd}=27,000$ K to 34,000 K for a distance of 200-250 pc. 
In that work, we did not considered non-solar abundances for SS Aur. 
These previous results \citep{god08,sio08} for the temperature and
gravity of SS Aur and TU Men are in line with our present global analysis, 
however, they only represent a small number of data points when
compared to Figs.\ref{tumenchi} \& \ref{ssaur_chi}. 
Overall, the present work comes to improve 
and enlarge our FUV spectral analysis of TU Men and SS Aur. 

We reconsider here our interpretation that the WD in SS Aur has solar
abundances, and show in the present work that data implies subsolar 
silicon abundance as well as possibly subsolar carbon abundance. 
Our most important result is that both TU Men and SS Aur display 
subsolar C and Si abundances, in addition
TU Men also presents evidence of suprasolar nitrogen abundance, 
i.e. evidence of CNO processing. 
We do not find
evidence of elevated nitrogen abundances for SS Aur. 
On the contrary, we find that the red wing of the N\,{\sc ii} $\lambda$ 1085
absorption feature (Fig.\ref{ssaurfusedetail}) 
agrees well with solar abundances, though the rest
of the feature is possibly contaminated with some geocoronal component.  

The STIS snapshots of SS Aur and TU Men have, respectively, an exposure time of 600 s and 900 s.
As a consequence the absorption lines are not broadened by the WD orbital motion. 
But the STIS snapshots have a low resolution which might affect some weak and 
sharp absorption lines, but cannot be responsible for the absence of C\,{\sc ii} (1175)
in SS Aur or the shallow Si\,{\sc ii+iii} (1300) absorption feature (unlike COS, 
STIS also does not suffer from much 
geocoronal emission O\,{\sc i} at 1300 \AA ). 

We found that both the FUSE and STIS spectra of SS Aur 
agree with a velocity broadening of 400 km/s 
within the limits of error bars,
with the lines in the FUSE spectrum being (on average) 
200~km/s broader than in the STIS spectrum. 
The FUSE spectrum of SS Aur was obtained over  a (raw) period of 
about 8~hr, or almost two complete binary orbits. 
The reason for the different broadening velocities in the FUSE
and STIS spectra is likely due to the combined effect of the 
WD orbital velocity ($K_1 = 70 \pm 10$ km/s, see Table \ref{syspar})
and the different instrumental broadening. 
We considered looking at the individual exposures (orbits) of the FUSE spectrum
of SS Aur, but their poor S/N (when not co-added) prevented us from being able
to properly fit the absorption lines.

For TU Men, $K_1$ could be twice as large, but since the STIS snapshot lasted only 
900 s, broadening due to the WD motion need not to be taken into account.   
Regardless of that, the STIS snapshots have a resolution of $R \sim 1000$,
with a binning of 0.58 \AA . At the STIS wavelengths, a $\Delta \lambda$ of 
0.58 \AA\ corresponds to a velocity of $\sim$100-150 km/s. 
For both objects, we can therefore take this velocity ($\sim 150$ km/s) 
as being the lower resolution limit for the measurement of line
broadening or measurement of the projected stellar rotational velocity.  
\\ \\ 

The high N/C abundance ratio implies CNO processing which elevates the 
abundance of nitrogen and depletes the abundance of carbon.
Interstingly, while the dwarf novae TU Men and VW Hyi reveal 
prominent C\,{\sc iv} emission line profiles in their FUV spectra during 
quiescence, their accreting white dwarfs show large N/C photospheric 
abundance ratios and subsolar C \citep{sio97, lon99}. The dwarf nova U Gem, 
which lies above the CV period gap, reveals no such C\,{\sc iv} emission line 
in either outburst or quiescence \citep{pan84}. Yet, the donor star 
in U Gem is carbon deficient while the donor star in 
VW Hyi appears to have solar carbon abundance \citep{har16}. 
The carbon deficiency in the photosphere of the white dwarf in VW Hydri
could be, at least in part, an effect of gravitational diffusion 
\citep[diffusion time scales of metals at the bottom of the convection zone 
in WDs can be as short as $\sim days$, see e.g.][for a short review and recent
developments]{hei20}.  
In our synthetic spectral  
modeling we do not include emission line modeling capability but it 
is virtually certain that this C\,{\sc iv} emission does not form on 
the accreting white dwarf itself. Instead, it may form in the inner disk 
or an accretion disk corona during quiescence when the accretion disk 
is optically thin.  
It cannot be entirely ruled out that the C\,{\sc iv} 
emission forms in the boundary layer between the inner disk and white dwarf. 
But even if the C abundance implied by the C\,{\sc iv}  emission line 
strength is solar, the presence of suprasolar N still points toward 
CNO processing. 

The exact origin of nonsolar abundances is not known, but several hypothesis
have been suggested \citep[see e.g.][for a short review]{gan03}.
For example the possibility of anomalous abundances in the 
accreting material itself, either from a (recent) nova outburst (where the
donor star is contaminated by the material ejected from the nova explosion
of the WD), or related to the nuclear evolution of the secondary
(if the donor was originally more massive than the WD, resulting in a 
transitional phase of unstable thermal timescale mass transfer,
it is then stripped from its outer layer). 
The secondary star in VW Hydri has a solar carbon abundance \citep{har16},  
yet the accreting white dwarf in VW Hydri has a large N/C abundance 
ratio as expected for hot CNO burning. While we do not know the carbon 
abundance of the donor secondary in TU Men, the accreting white dwarf 
photosphere also has a suprasolar N/C abundance ratio. This result 
quite possibly could be pointing toward the origin of the elevated N 
and depleted C in the white dwarf itself \citep{sio14}. This may support a scenario 
whereby the elevated N/C ratio does not arise from the peeling away 
by mass transfer of the CNO processed core of an originally more massive 
donor secondary but instead in the WD itself from the hot CNO processing 
of many previous nova explosions \citep{sio98, sio14}. 
It is important to remark also that, while other CVs have exhibited an anomalously 
high N/C ratio, TU Men is the only dwarf nova with an exposed WD in the 
CV period gap.

As to SS Aur, the lack of strong carbon absorption lines in the FUSE and STIS spectra
strongly contrasts with the  presence of normal looking CO absorption
in the K-band from its secondary \citep{how10}, and C\,{\sc iv} 
emission line (see Fig.\ref{ssaurstis}) from disk, disk corona or boundary layer. 

Despite not having phase-resolved high S/N spectra (available
only for a few of systems, such as VW Hyi and U Gem), we have presented 
further evidence of subsolar C abundances in two accreting white
dwarfs, in the dwarf novae TU Men and SS Aur.  
As part of our HST Archival Program research (see Acknowledgements
below), we are currently carrying out chemical abundance studies 
in more systems with exposed accreting white dwarfs in order
to further understand the physics of white dwarf accretion and  
the evolution of cataclysmic variables.

\acknowledgements 
Support for this research is provided by NASA through grant number 
HST-AR-16152 to Villanova University from the Space Telescope Science
Institute, which is operated by AURA, Inc., under NASA contract
NAS 5-26555.  
PG is pleased to thank William (Bill) P. Blair at the 
Henry Augustus Rowland Department of Physics \& Astronomy at The 
Johns Hopkins University, Baltimore, Maryland, USA, for his 
indefatigable kind hospitality. 

\software{
IRAF \citep[NOAO PC-IRAF Revision 2.12.2-EXPORT SUN;][]{tod93}, 
Tlusty (v203) Synspec (v48) Rotin(v4) \citep{hub17a,hub17b,hub17c}, 
PGPLOT (v5.2), Cygwin-X (Cygwin v1.7.16),
xmgrace (Grace v2), XV (v3.10) } 
\\ \\

\begin{center} 
{\bf{ ORCID iDs}} 
\end{center} 
Patrick Godon \url{https://orcid.org/0000-0002-4806-5319}  \\ 
Edward M. Sion \url{https://orcid.org/0000-0003-4440-0551}

\end{document}